\documentclass[preprint,12pt]{elsarticle}
\usepackage{graphicx} 
\usepackage{lineno}
\usepackage[latin9]{inputenc}
\usepackage{hyperref}
\usepackage{amsmath}
\usepackage{dsfont} 
\usepackage{multirow} 
\usepackage{booktabs}
\usepackage{amssymb}
\usepackage{subfig}
\usepackage{rotating} 
\usepackage{xcolor}

\usepackage[acronym, toc, nopostdot, indexonlyfirst,numberedsection=autolabel,automake]{glossaries}
\usepackage{glossary-mcols} 
\glsdisablehyper
\makeglossaries
\newacronym{ECG}{ECG}{electrocardiogram}
\newacronym{PPG}{PPG}{photoplethysmogram}
\newacronym{CHF}{CHF}{Congestive Heart Failure}
\newacronym{HF}{HF}{Heart Failure}
\newacronym{HRV}{HRV}{Heart Rate Variability}
\newacronym{ML}{ML}{machine learning}
\newacronym{DL}{DL}{deep learning}
\newacronym{RRI}{\textit{RRI}}{\textit{RR} intervals}
\newacronym{SE}{SE}{Squeeze-and-Excitation}
\newacronym{HR}{HR}{Heart Rate}
\newacronym{CPH}{CPH}{Cox proportional hazards}
\newacronym{AFT}{AFT}{Accelerated Failure Time}
\newacronym{RSF}{RSF}{Random Survival Forest}
\newacronym{EST}{EST}{Extremely randomized Survival Trees}

\journal{Information Fusion}

\begin{document}

\begin{frontmatter}

\title{Multi-modal Heart Failure Risk Estimation based on Short ECG and Sampled Long-Term HRV}

\author[aic]{Sergio~Gonz\'alez\corref{cor1}}
\ead{gonzalez-vazquez.sergio@inventec.com}
\cortext[cor1]{Corresponding author: Sergio~Gonz\'alez}
\author[aic]{Abel~Ko-Chun~Yi}
\author[aic]{Wan-Ting~Hsieh}
\author[aic]{Wei-Chao~Chen}
\author[cd,cm]{Chun-Li~Wang}
\author[cd,cm]{Victor~Chien-Chia~Wu}
\author[cd]{Shang-Hung~Chang}

\myfntext[license]{2024. Licensed under the \href{https://creativecommons.org/licenses/by-nc-nd/4.0/}{CC-BY-NC-ND 4.0 license}.}

\address[aic]{AI Center, Inventec Corporation, Taipei, Taiwan}
\address[cd]{Cardiovascular Division, Department of Internal Medicine, Chang Gung Memorial Hospital, Linkou, Taiwan}
\address[cm]{College of Medicine, Chang Gung University, Taoyuan, Taiwan}

\begin{abstract}

Cardiovascular diseases, including Heart Failure (HF), remain a leading global cause of mortality, often evading early detection. In this context, accessible and effective risk assessment is indispensable. Traditional approaches rely on resource-intensive diagnostic tests, typically administered after the onset of symptoms. The widespread availability of electrocardiogram (ECG) technology and the power of Machine Learning are emerging as viable alternatives within smart healthcare.
In this paper, we propose several multi-modal approaches that combine 30-second ECG recordings and approximate long-term Heart Rate Variability (HRV) data to estimate the risk of HF hospitalization. We introduce two survival models: {an XGBoost model with Accelerated Failure Time (AFT)} incorporating comprehensive ECG features and a ResNet model that learns from the raw ECG. We extend these with our novel long-term HRVs extracted from the combination of ultra-short-term beat-to-beat measurements taken over the day. To capture their temporal dynamics, we propose a survival model comprising ResNet and Transformer architectures (TFM-ResNet). {Our experiments demonstrate high model performance for HF risk assessment with a concordance index of 0.8537 compared to 14 survival models and competitive discrimination power on various external ECG datasets.} After transferability tests with Apple Watch data, our approach implemented in the \textit{myHeartScore} App offers cost-effective and highly accessible HF risk assessment, contributing to its prevention and management.

\end{abstract}


\end{frontmatter}

\section{Introduction} \label{sec:introduction}

Cardiovascular diseases are one of the leading causes of death in many countries around the world. In many cases, these diseases silently develop without clear symptoms until the patient's health is compromised. Among them, \gls{HF} is a highly lethal medical condition caused by cardiac dysfunction \cite{riccardi2022heart}. In developed countries, a significant portion of the healthcare budget is allocated to managing \gls{HF} \cite{savarese2017global}. Given these circumstances, risk assessment and prevention play a crucial role in preserving both lives and resources.

Traditional methods of risk assessment of \gls{HF} rely on a combination of patient history, clinical examination, and diagnostic tests \cite{kannel1999profile,butler2008incident,agarwal2012prediction}. However, these methods can be inaccessible, costly, and require significant effort. Consequently, they are typically employed when patients exhibit symptoms or clear risk factors. {In addition, they use classical survival analysis models like \gls{CPH} regression \cite{kleinbaum2010survival}, which may make unsuitable assumptions for the problem \cite{rulli2018assessment}.} To address these limitations, ongoing efforts are focused on developing more accessible and cost-effective approaches \cite{wang2019machine,olsen2020clinical}.

One such approach involves utilizing the \gls{ECG} \cite{dukes2015ventricular,patel2017association}, a highly accessible and affordable technology capable of monitoring heart activity. Additionally, \gls{ML} techniques are emerging as a viable alternative \cite{wang2019machine,olsen2020clinical}. Recent \gls{ML} approaches based on short \gls{ECG} strips have demonstrated promising results \cite{akbilgic2021ecg,gonzalez2022interpretable}. However, they might lack a comprehensive understanding of heart activity throughout the day, potentially missing relevant sporadic cardiac abnormalities. Conversely, collecting long-term ECG recordings may be less practical due to the requirement for specialized and cumbersome devices.

Multi-modal learning, which has proven effective in smart healthcare \cite{shaik2023survey, niu2024ehr}, leads the way to overcoming these restraints. {In addition, \gls{HRV} statistics, which measure the physiological phenomenon of temporal variation between heartbeats, have been used clinically with long-term measurements ($\ge$ 24 hours) \cite{electrophysiology1996heart}. However, ultra-short ($\le$ 5 min) HRVs are growing in popularity thanks to advances in \gls{PPG} and wearable technologies and are associated with long-term risks of cardiovascular events \cite{orini2023long}. Therefore, HRVs combined with short ECG recordings may be a practical alternative with easy deployment in existing wearable devices, such as smartwatches.} 

{Hence, we propose a multi-modal approach that learns from a 30-second ECG recording, approximate long-term \gls{HRV} from sampled ultra-short beat-to-beat measurements, demographic information, and the basic medical history of the patient to estimate the risk of HF hospitalization.} We have designed two survival models: an XGBoost model with \gls{AFT} \cite{barnwal2020survival} with comprehensive features of the ECG rhythm and PQRST complex, and a ResNet model that adapts DeepHit loss \cite{lee2018deephit} to learn the risk factors directly from the ECG signal.

Besides, we propose a novel procedure to extract approximate long-term \gls{HRV} statistics through the combination of short beat-to-beat measurements sampled over 24 hours, {which can be extracted from other modalities, such as \gls{PPG} \cite{lu2009comparison, jeyhani2015comparison}, with common wearable devices.} These HRVs can be incorporated into our models as aggregated independent features or as HRV time series to include a more complete understanding of the heart activity and boost model performance. We have extended our ResNet model with a Transformer (TFM) module to exploit the time dependency of our HRV series.

Our proposals can enable an accessible, non-invasive, and cost-effective method for assessing the risk of HF. They are easy to implement on any wearable or smartwatch, making them highly accessible and convenient. To demonstrate it, we present \textit{myHeartScore} App as one of its use cases. This iPhone App leverages the ECG and beat-to-beat measurements of the Apple Watch and our ML proposals to provide insights into the user's health including our risk score. Users can periodically reassess their scores, potentially enabling the early detection of cardiac abnormalities and the prevention and management of HF.

To test the proposed methods, we perform several experiments on their performance in survival settings and other external ECG datasets. {First, we test 15 different survival models with a held-out test set of 6,344 subjects, demonstrating an improved performance by our multi-modal models. Then, we study the effect of varying amounts of beat-to-beat samples on our approximate long-term HRV, showing similar model performance to that of the exact 24-hour HRV.} Next, we analyze the discriminative power of our models with other open ECG datasets for HF classification: MIMIC-III \cite{johnson2016mimic}, BIDMC \cite{baim1986survival}, and NSRDB \cite{goldberger2000physiobank}. Our TFM-ResNet model performs similarly to other HF classification algorithms \cite{acharya2019deep, kusuma2022ecg}. Finally, we test the transferability of our models with self-collected data compromising 89 individuals and 413 ECG recordings from the Apple Watch. {Thus, we demonstrate that our multi-modal approaches with short ECG and sampled long-term HRV are a highly accessible and convenient alternative for HF risk assessment.}

This paper is organized as follows. In Section \ref{sec:background} we present survival analysis and the related works of HF risk estimation. Section \ref{sec:proposal} is devoted to our HF risk estimator models with the novel procedure of extracting the sampled long-term HRVs. In Section \ref{sec:experiments} we describe the experimental framework followed by the empirical results and the analysis of our models' performance and transferability to other ECG datasets. Section \ref{sec:myheartscore} introduces \textcolor{black}{myHeartScore App \cite{myheartscore}} as a use case to exploit the potential of our predictive models. Finally, Section \ref{sec:conclusions} closes the paper with its main conclusions.

\section{{Related work}}
\label{sec:background}

{Survival analysis studies the expected time until an event of interest \cite{kleinbaum2010survival}, such as the death in a group of patients or a disease occurrence in a population. In these survival settings, the outcome $\delta^i \in {1,0}$ denotes whether the event occurred during $y_i$, the time-to-event or observation time of subject $i$. This notation accounts for the frequent existence of censored data, i.e., individuals with partial information about their survival times. For example, right-censored individuals survived during their observation time, but whether the event can occur afterward is unknown. Survival models are particular regression models capable of learning from time-to-event and censoring data.}

{\gls{CPH} and \gls{AFT} are popular classic survival models due to their simplicity and straightforward interpretability \cite{kleinbaum2010survival}. However, they are limited as linear models and make strong assumptions about the survival problem \cite{rulli2018assessment}. For instance, they do not consider the interaction between features with constant effects over time, their performance deteriorate with high dimensional or correlated data, and they cannot directly learn from non-tabular data, such as images or signals. Therefore, machine learning and deep learning survival approaches are getting more attention lately \cite{wang2019machine}. Some well-known ML survival models are \gls{RSF} \cite{10.1214/09-SS047}, \gls{EST} \cite{10.1214/09-SS047}, and XGBoost \cite{chen2016xgboost}, the flagship of the ensemble algorithms \cite{gonzalez2020practical} that extends \gls{CPH} and \gls{AFT} in its gradient boosting engine \cite{barnwal2020survival}.} DeepSurv \cite{katzman2018deepsurv}, DeepHit \cite{lee2018deephit}, and Cox-Time \cite{kvamme2019time} are popular deep learning approaches.

{HF risk estimation has traditionally relied on the \gls{CPH} model combining information from demographics, medical history, costly laboratory, and in-hospital tests \cite{kannel1999profile,butler2008incident,agarwal2012prediction}. Besides, other statistical studies \cite{dukes2015ventricular,patel2017association} have demonstrated the benefits of long-term ECG recording parameters, such as \gls{HRV} and premature ventricle contraction frequency, when estimating \gls{HF} risk.} 

{More recently, machine learning-based approaches have been developed to ease \gls{HF} classification, prognosis, and prediction \cite{olsen2020clinical}. Deep learning methods, such as CNN and LSTM, have been successfully adapted for HF diagnosis using ECG signals \cite{acharya2019deep, kusuma2022ecg}. \gls{ML}-based survival models were also used to estimate the mortality risk in heart failure patients \cite{li2021prediction, yang2021application}.} For \gls{HF} risk estimation, several \gls{ML} studies have shown improvements compared to traditional risk calculators \cite{akbilgic2021ecg,segar2019machine,segar2021development,zhou2023risk}. However, the majority still rely on costly laboratory and diagnostic tests. Only a few \textcolor{black}{studies} \cite{akbilgic2021ecg, gonzalez2022interpretable} use short ECG measurements as their risk factor. One of these \cite{akbilgic2021ecg} implemented a ResNet model with 10s 12 lead ECGs to predict the development of HF within ten years in a classification fashion. Then, its output was combined with traditional risk factors to build a \gls{CPH} model. Our approach differs, as our survival models derive the survival outcome directly from the ECG. In addition, we propose a multi-modal approach to include additional information from cardiac activity throughout the day.  

{Our previous study \cite{gonzalez2022interpretable} proposed an interpretable HF risk estimator based on XGBoost. However, our model was not tested with external data and the PQRST features lack the original ECG units. In this paper, we improve these aspects and propose a novel multi-modal approach model that learns directly from short ECG signals and approximate long-term HRVs.}

\section{Multi-modal HF risk estimation based on short ECG and sampled long-term HRV}
\label{sec:proposal}


In this section, we explain our multi-modal models for estimating the risk of \gls{HF} hospitalization using a short 30-second ECG signal, sampled long-term HRVs, and the previous conditions of the patient. {As aforementioned, we have developed two different learning models: an XGBoost AFT with comprehensive ECG features detailed in Subsection \ref{subsec:feat-model}, and a ResNet model that learns directly from the raw ECG signal, as presented in Subsection \ref{subsec:raw-model}.} Finally, we introduce our extraction of approximate long-term \gls{HRV} statistics by combining different ultra-short beat-to-beat measurements taken over 24 hours, and how they are incorporated into the previous models to boost their performance.


\subsection{{XGBoost AFT survival model based on comprehensive ECG features}}
\label{subsec:feat-model}

Our feature-based model requires the preprocessing and the extraction of meaningful features of the 30s ECG signal. Prior to any procedure, we transform our Holter 3-channel ECG data to a standard 12-lead ECG data using the EASI linear coefficients for derived 12-lead ECG \cite{feild2002improved}. This transformation enables us to use the lead I for our models, making them more general and applicable to other data or devices, such as smartwatches. 

\begin{figure}[ht] 
\centering 
\includegraphics[width=\textwidth]{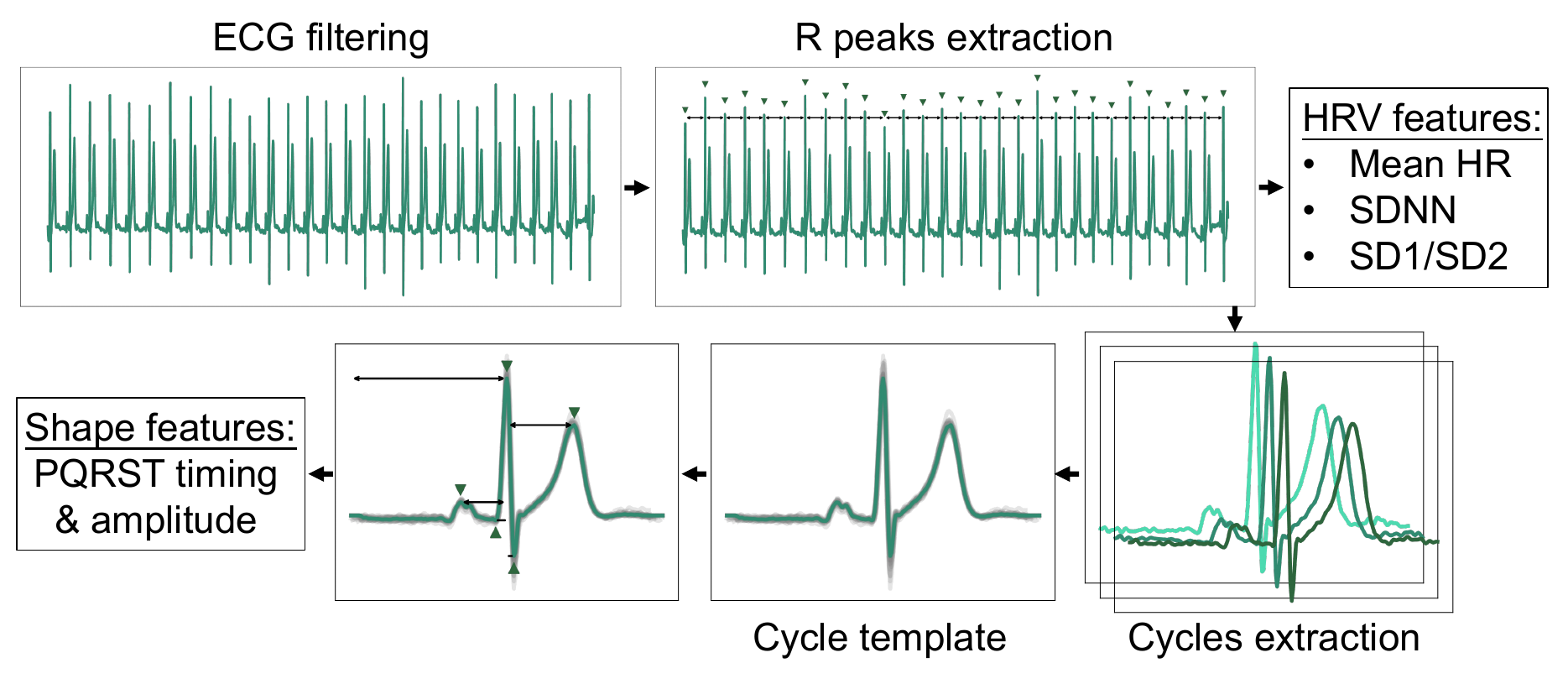} 
\caption{Preprocessing and features extraction of a 30s ECG. } 
\label{fig:feats2}
\end{figure}

Next, we proceed with the preprocessing and feature extraction shown in Figure \ref{fig:feats2}.
First, we filter the 30s lead I ECG signal with a Butterworth band-pass filter 
between 0.05 and 45 Hz to mitigate device noise and movement artifacts. We identify the \textit{R} peaks of the ECG using Hamilton's algorithm \cite{hamilton1986quantitative} and compute the \gls{RRI} as the time difference between consecutive \textit{R} peaks. We extract all cardiac cycles with their \textit{R} peak centered and length equal to the shortest distance to the adjacent \textit{R} peaks. We add some padding so that all cycles have the same size, align them on their \textit{R} peaks, and aggregate with the median. The result is a cycle template with preserved timing and amplitude of the \textit{PQRST} complex.

With the \gls{RRI}s and the cycle template, we extract two types of interpretable ECG features:
\begin{itemize}
    \item \textbf{Short HRV features based on the \gls{RRI}s:}
    \begin{itemize}
        \item \textit{Mean HR}: the mean heart rate measured as beats per minute.
        \item \textit{SDNN}: The standard deviation of the RR intervals measured in milliseconds.
        \item \textit{SD1/SD2}: The ratio of the two standard deviations of the Poincar\'e analysis \cite{golinska2013poincare} defined as $RRI[i]$ and 
        $RRI[i+1]$ in the x and y axes, respectively. 

\end{itemize}
    
    \item \textbf{Shape features characterizing the ECG cycle template:} The \textit{P}, \textit{Q}, \textit{S}, and \textit{T} waves are identified as the most prominent consecutive peaks and valleys of the cardiac cycle. \textit{P} and \textit{T} inversions are considered possible cases in our delineation procedure. The timing of each wave is computed as the time difference to the R peak. Its amplitude is calculated as the deviation from the baseline, which is defined as the median of the template.
\end{itemize}

These ECG features are paired with the demographics and known medical conditions of the subject. The numeric features are normalized using a quantile transformation, which alleviates potential issues due to outliers and skewed distributions. 

\begin{figure}[ht] 
\centering 
\includegraphics[width=\textwidth]{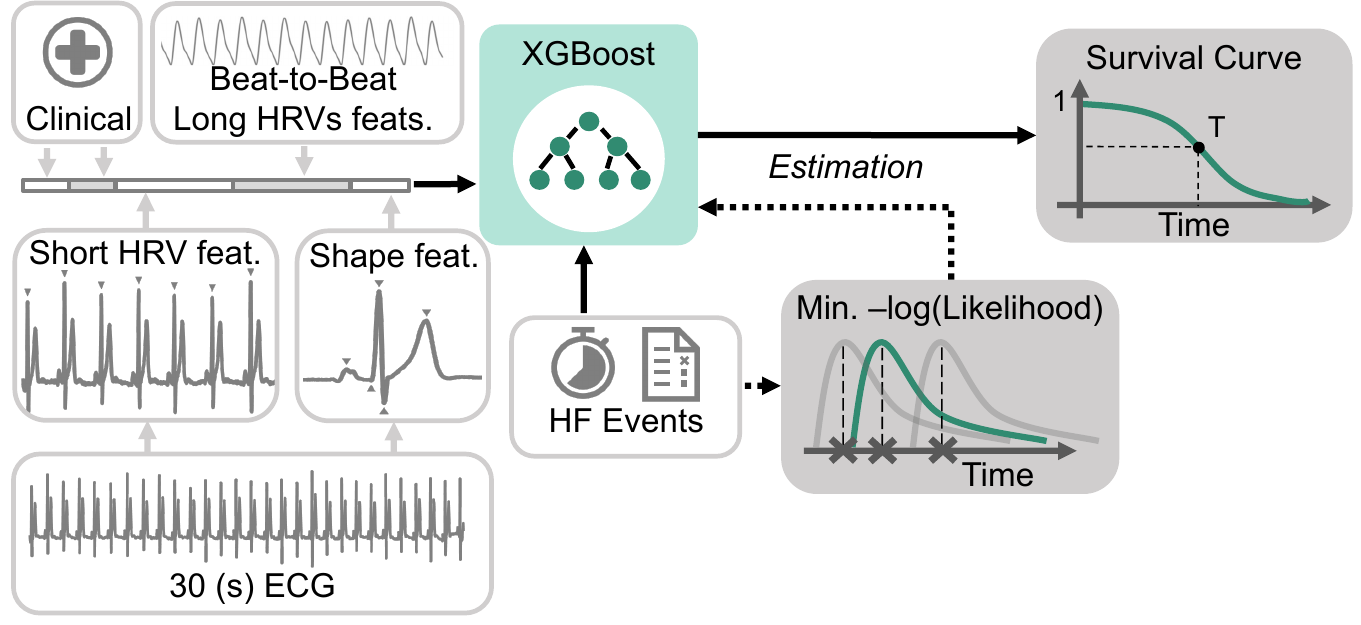} 
\caption{Our features-based model relies on XGBoost AFT and the extraction of relevant features from the 30s ECG recording and the beat-to-beat measurements.} 
\label{fig:xgb}
\end{figure}

Our feature extraction enables us to use any \gls{ML} survival analysis model to estimate the \gls{HF} risk. We chose XGBoost AFT due to its effectiveness and higher performance over other models in our previous study \cite{gonzalez2022interpretable}. XGBoost AFT \cite{barnwal2020survival} is an effective gradient boosting algorithm that generalizes the ``accelerated failure time" \cite{kleinbaum2010survival} for survival settings. The model operates under the assumption that $ln(T)$ follows the equation $ln(T) = \tau(\mathbf{x}) + \epsilon$. Here, $T$ represents the time-to-event random variable, $\tau(\mathbf{x})$ is the estimation of the gradient boosting engine, and $\epsilon$ is the error term. We are using a log-logistic (Fisk) distribution for our \gls{HF} time-to-event distribution. That is, the error term $\epsilon$ conforms to a logistic distribution with a zero mean and a standard deviation, denoted as the hyperparameter $\sigma$. Consequently, we can readily convert the predicted times of XGBoost into survival curves. To train the model, the negative log-likelihood loss function is used. Given the imbalanced nature of the problem, we implement instance weighting and introduce L1 and L2 regularization to enhance model performance and stability. {Figure \ref{fig:xgb} represents the learning procedure of our features-based XGBoost AFT model, which includes only clinical and ECG-based features in this version.}

Finally, it's worth highlighting that our feature-based model, while potentially exhibiting a lower performance compared to the raw signal-based model, has the advantage of being easier to interpret and lighter allowing it to be implemented directly on device.

\subsection{{ResNet survival model based on raw ECG signal}}
\label{subsec:raw-model}

\begin{figure}[ht] 
\centering 
\includegraphics[width=\textwidth]{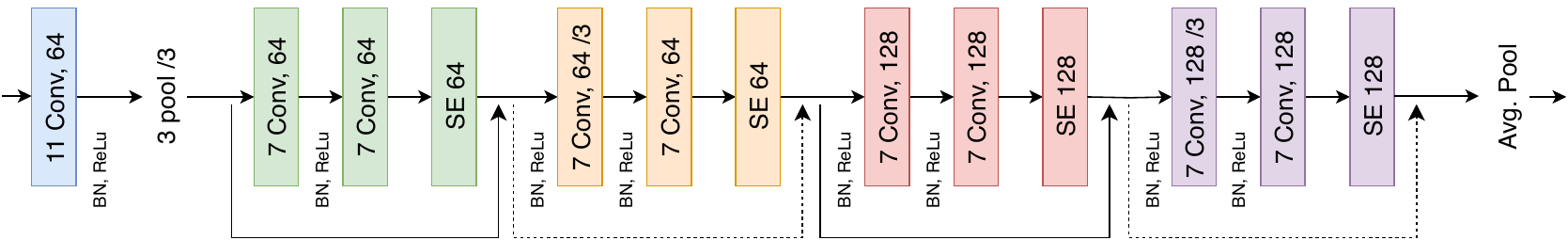} 
\caption{ResNet architecture with 4 residual blocks. In each convolutional layer, we indicate the kernel size, number of filters, and stride size, whose default value is 1. For example, 7 Conv. 64 /3 represents a 1D convolutional layer with a kernel size equal to 7, 64 filters, and a stride size of 3.} 
\label{fig:rsnt-arch}
\end{figure}

The architecture of our raw signal-based model has two distinguishable parts: the automatic feature extractor and the discriminator. The automatic feature extractor infers valuable latent features from the lead I 30s raw \gls{ECG}. To do so, we have designed a ResNet architecture \cite{he2016deep} with 1D CNN. Figure \ref{fig:rsnt-arch} shows a representation of our ResNet architecture, which has one initial extraction layer and 4 residual blocks. Each residual block consists of two layers of 1D CNN, batch normalization, and ReLU activation followed by a \gls{SE} block \cite{hu2018squeeze}. \gls{SE} blocks help to adjust the importance of different feature channels based on their interdependencies. The squeezing consists of an adaptive average pooling. The excitation is implemented with a linear two-layer bottleneck with a ReLU activation in between and a final sigmoid activation. Finally, the outputs of our ResNet are averaged in the dimension of the features,  concatenated with normalized subject information, and passed to the discriminator.

\begin{figure}[ht] 
\centering 
\includegraphics[width=\textwidth]{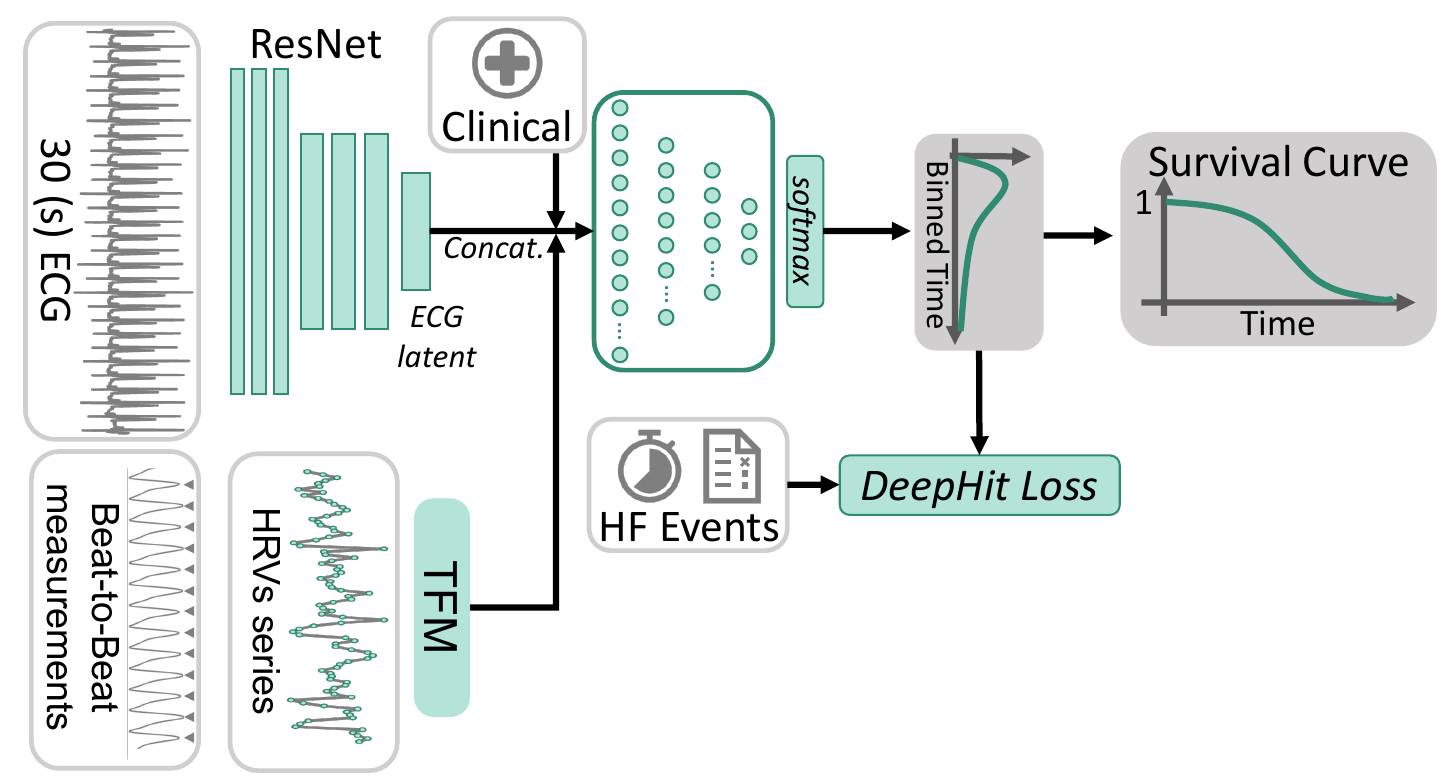} 
\caption{TFM-ResNet model includes a ResNet model to extract learned features from the 30s ECG signal, and a Transformer model encodes the HRVs time series. Then, an MLP combines these two sources of information and personal information to output the survival curve.} 
\label{fig:rsnt}
\end{figure}

Our discriminator has four stacked linear layers of 32 neurons, LeakyReLU activation function, and a dropout of 0.1. The final layer has 26 outputs that represent the probability distribution over a grid of discrete-time steps, that is $f(\mathbf{x}) = [p(t_1),...,p(t_n)]^T$. Thus, we divided the observation time of \gls{HF} events, about 11 years, into 25 equidistant time intervals. The remaining output corresponds to the probability of not developing \gls{HF} within the observation time. {Figure \ref{fig:rsnt} shows an overview of our raw signal-based survival model, which excludes the Transformer architecture for the model with only ECG.}

We have trained our model in a survival setting with an adapted version of the DeepHit loss function \cite{lee2018deephit}. The original DeepHit loss is a weighted average of the log-likelihood and rank loss terms, i.e. $\mathcal{L}=\alpha \mathcal{L}_1+(1-\alpha)\mathcal{L}_2$. $\mathcal{L}_1$ is the log-likelihood of the predicted probability distribution and the time-to-event adapted to right-censored data. For censored data, it exploits the observed time by taking the logarithm of one minus the cumulative probability during that time. To address the substantial data imbalance, we have modified this term to incorporate Focal Loss \cite{lin2017focal}:

\begin{equation}
\begin{split}
    \mathcal{L}_1 = - &\sum_{i=0}^N[\delta^i \cdot (1-P_i(t=y_i))^\gamma \cdot \log P_i(t=y_i) \\ 
    &+ (1-\delta^i) \cdot (1-P_i(t \leq y_i))^\gamma \cdot \log P_i(t \leq y_i)]
\end{split}
\label{eq:nll}
\end{equation}

\noindent where $\delta^i$ takes 1 if the event has occurred for subject $i$ or 0 otherwise, $y_i$ is the time-to-event or observation time, and $\gamma$ is the Focal Loss factor. The rank loss $\mathcal{L}_2$ incorporates the concept of the concordance index \cite{harrell1982evaluating}, that is individuals who experience the event earlier should have a higher risk: 

\begin{equation}
    \mathcal{L}_2 = \sum_{i \neq j } \mathds{I}(\delta^i, y_i < y_j) \exp{(\frac{-(P_i(t \leq y_i)-P_j(t \leq y_i))}{\sigma})}\\
\label{eq:rankloss}
\end{equation}

\noindent where $\mathds{I}$ is an indicator function for subject $i$ experiencing the event ($\delta^i$), and its time to event $y_i$ is shorter than the time $y_j$ of the other subject $j$.

Finally, we have integrated data augmentation into our model training. Specifically, we randomly scaled 50\% of the batch by a factor ranging from 0.8 to 1.2. This strategy serves to prevent overfitting and enhance the model's generalization capabilities.

\subsection{{Approximate long-term HRVs from sampled beat-to-beat measurements}}
\label{subsec:hrvs}

The models explained above rely only on a 30s ECG signal, which might have the drawback of missing some infrequent cardiac abnormalities or lacking a complete understating of the heart activity. Therefore, we have extended our models to incorporate approximate long-term \gls{HRV}s from sampled beat-to-beat measurements that mitigate these issues. 

Our long-term \gls{HRV}s from sampled beat-to-beat measurements are heart-rate-related statistics computed with the combination of several ultra-short samples (between 1 to 5 minutes) collected during a long period, such as 24 hours. They can be seen as an approximation of the actual \gls{HRV}s of that period. {This sampling approach reduces the computational cost of data recording and preprocessing while maintaining a good understanding of the heart activity during the day. Existing wearable devices, such as smartwatches, only measure a few beat-to-beat measurements a day because users may not wear them all day, some measurements may be discarded due to noise or motion artifacts, and the devices might prioritize battery life. Thus, our approach enables a seamless implementation in smartwatches, such as Apple Watch.} 

\begin{figure}[ht] 
\centering 
\includegraphics[width=\textwidth]{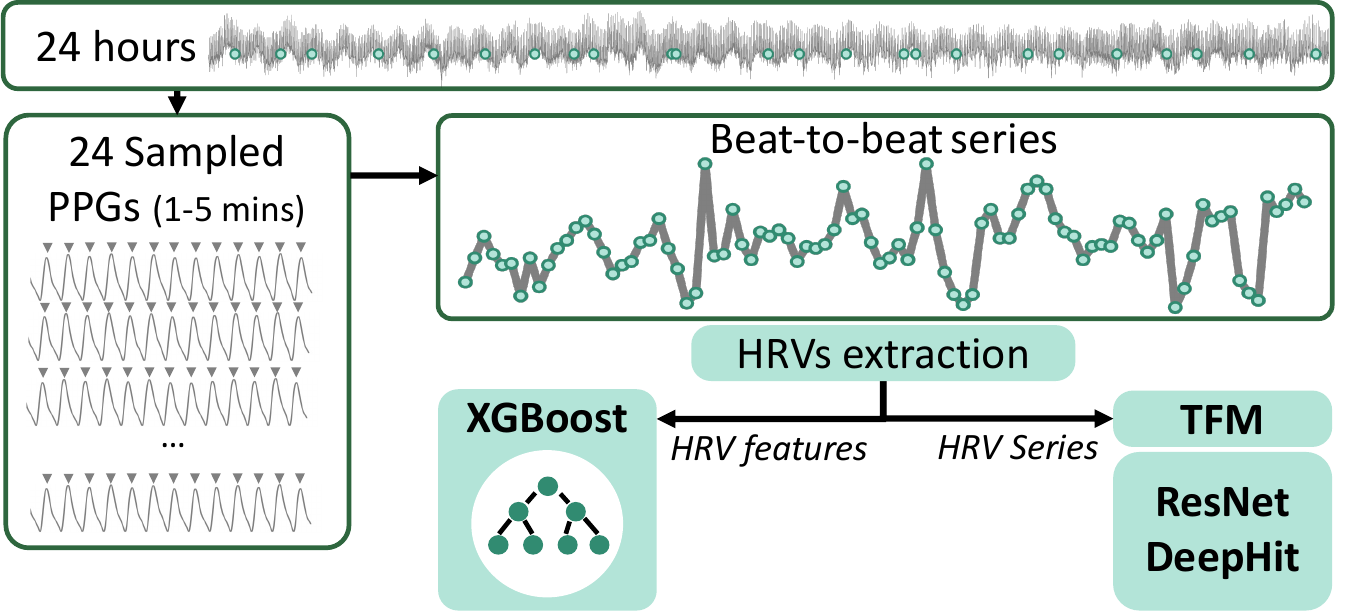} 
\caption{Flowchart of the implementation of the long-term HRVs from sampled beat-to-beat measurements.} 
\label{fig:long-hrvs}
\end{figure}

Figure \ref{fig:long-hrvs} exemplifies the implementation of our sampled long-term HRVs into our models. Our training dataset only has 24-hour ECG recordings from Holter machines. Thus, we extract the \gls{HRV}s from the \glsfirst{RRI} of \gls{ECG} samples. {But, the same beat-to-beat information can be obtained from simpler \gls{PPG} waveforms of wearable devices with marginal differences \cite{lu2009comparison, jeyhani2015comparison}.}

From the 24-hour recording, we randomly sample at least one ultra-short strip (1-5 mins.) per hour. We filter the signals as aforementioned and identify beat-to-beat measurements, i.e. the time passed between consecutive heartbeats. This process results in 24 beat-to-beat series. Then, we computed relevant statistics of the \gls{HR} and \gls{HRV}. To do so, we distinguish two different procedures depending on the implemented model: 

\begin{itemize}
    \item {\textbf{Single HRV features vector approach:}} This procedure first concatenates the beat-to-beat measurements into one single series. Then, the HRV statistics are computed from the whole series, avoiding nonconsecutive beat-to-beat intervals. The output is a single vector of sampled long-term HRV features, which can be concatenated with the ECG features and the personal information and fed to any feature-based model, such as our XGBoost (Figure \ref{fig:xgb}) or the MLP of our ResNet model. Our HRV features are \textit{active \gls{HR}}, \textit{\gls{HR} at rest}, \textit{SDNN}, \textit{RMSSD}, \textit{PNN50}, \textit{Sample Entropy}, \textit{SD1/SD2}, and the \textit{ellipse's area} of the Poincar\'e plot.
    \begin{itemize}
        \item Active \gls{HR}: 85th percentile of the \gls{HR} of the 24 sampled beat-to-beat measurements.

        \item \gls{HR} at rest: 15th percentile of the \gls{HR} of the 24 sampled beat-to-beat measurements.
        
        \item \textit{RMSSD}: Root mean of the squared beat-to-beat measurement differences ($\Delta BB$).
        \begin{equation}
            \label{eq:rmssd}
            \text{\textit{RMSSD}} = \sqrt{\frac{1}{K-1}\sum_{i=1}^{K}\Delta BB_i^2} 
        \end{equation}

        \item \textit{PNN50}: Ratio of consecutive \textit{BB} differences greater than 50 milliseconds.

        \item \textit{Sample Entropy}: The sample entropy of the beat-to-beat measurements is defined as the negative logarithm of the conditional probability that two similar subsequences (each consisting of 2 points) remain similar at the following point. Similarity is determined based on whether the difference between the subsequences is less than the \textit{SDNN}.
    \end{itemize}
    
    \item {\textbf{HRV time series approach:} We compute first the HRV statistics as aforementioned, but for each set of beat-to-beat measurements independently, leading to an HRV time series.} Thus, the series can be used in any time series or sequence model, such as GRU \cite{cho2014learning} or Transformer (TFM) \cite{vaswani2017attention}. In our TFM-ResNet shown in Figure \ref{fig:rsnt}, we have designed a TFM with two encoding layers consisting of a self-attention block, addition \& normalization operations, a feed-forward block with a hidden dimension of 64, and a final addition \& normalization. Thus, the HRV series as input of our TFM comprises 24 samples, one per hour, of the previously mentioned HRV statistics. The outputs of our TFM model are averaged into a single features vector and concatenated with the ResNet's latent features and the subject information. Note that the Active \gls{HR} and \gls{HR} at rest are still computed with the merged beat-to-beat measurement and fed into the MLP directly.

\end{itemize}

\section{Experimental Framework, Results and Analysis}
\label{sec:experiments}

In this section, we present the experimental study carried out with our \gls{HF} risk estimation models. In Subsection \ref{subs:experimentalframe}, we introduce the experiment framework, noting the datasets, and evaluation metrics used. Subsection \ref{subs:exp-performance} is dedicated to analyzing the performance of our models in a survival analysis setting. {Next, we study the implication of varying beat-to-beat information on our approximate long-term HRV in Subsection \ref{subs:exp-hrv}.} In Subsection \ref{subs:exp-opendata}, we demonstrate the discriminative power of our models with other open ECG datasets. Finally, Subsection \ref{subs:exp-watch} shows our models' behavior with data self-collected using the Apple Watch ECG app. 

\subsection{Experimental framework}
\label{subs:experimentalframe}

\begin{table}[ht]
\resizebox{\textwidth}{!}{
    \centering
    \begin{tabular}{l|c|c|c|c|l}
    \toprule
         \textbf{Dataset}&  \begin{tabular}[c]{@{}c@{}}\textbf{Recording} \\ \textbf{Duration}\end{tabular} &  \textbf{Subjects}&  \textbf{Samples (30s)}&  \begin{tabular}[c]{@{}c@{}}\textbf{Sampling} \\ \textbf{rate (Hz)}\end{tabular} & \textbf{Outcome}
\\ \midrule
         Holter ECG& 24 hours &  21,891&  21,891&  200&  \begin{tabular}[c]{@{}l@{}}HF within 11 yrs: 8.2\% \\ Avg. Time-to-HF: 1.85 years\end{tabular} 
\\ \midrule
         MIMIC-III \cite{johnson2016mimic}&  $>$ 6 hours&  2,201&  2,201&  125& HF (9.6\%) / No (90.4\%)
\\ \midrule
         HF classification& - &  33&  3,300& -  & HF (45.5\%) / No (54.5\%)
\\ 
         \hspace{3mm} BIDMC \cite{baim1986survival}&  20 hours&  15&  1,500&  250& HF (NYHA class 3-4)
\\ 
         \hspace{3mm} NSRDB \cite{goldberger2000physiobank}&  $\thicksim$24 hours &  18&  1,800&  128& Healthy
\\ \midrule
         Apple Watch &  30 seconds&  89&  413&  512& 
\begin{tabular}[c]{@{}l@{}}{\underline{Apple results:} SR: 80.1\%,} \\ 
High HR: 4.1\%, AFib: 2.1\% \\ 
\underline{Risk factors:} No: 89.5\%, \\
{HTN: 6.1\%, Heart prob.: 4.4\%} \\ 
\end{tabular}\\
         \bottomrule
    \end{tabular}}
    \caption{Summary of the four datasets used in this study.}
    \label{tab:datasets}
\end{table}

In our experiments, we have used four different datasets to test our model performance as shown in Tables \ref{tab:datasets} and \ref{tab:datasets-feats}:  

\begin{itemize}
    \item \textbf{Holter ECG data:} were gathered by the Chang Gung Memorial Hospital in Linkou, Taiwan using three-lead Holter machines for 24 hours from 01/05/2009 to 12/29/2017. The cohort consisted of 21,891 individuals, with an even gender distribution of 50.4\% men and 49.6\% women. Patients were monitored for HF until 2020 for a minimum of two years. During the monitoring period, 1,805 patients (8.2\%) were admitted to the hospital due to \gls{HF}, with 919 (4.2\%) within the first year and 1,195 (5.5\%) within the first two years. The diagnosis of HF was established as their hospitalization with a primary discharge diagnosis of HF. We considered March 6, 2020, as the censoring date for patients who had not experienced HF by that time. Note that we did not consider death or other clinical events as censoring factors.

    \item \textbf{MIMIC-III \cite{johnson2016mimic}}: is a well-known dataset that includes different physiological signals and medical records of patients of the intensive care units (ICU). Our MIMIC III subset consists of 2,201 patients with sufficiently long lead I ECG measurements, of whom approximately 9.6\% had a HF diagnosis. We have selected a subgroup of HF patients previously used for mortality prediction \cite{li2021prediction} and added other individuals with diagnosis ICD codes different from those representing HF. We have excluded those subjects without lead I ECG recordings. 

    The characteristics of the MIMIC III dataset are different from our Holter ECG set, because the former patients are in ICU, while the latter are outpatients. This may affect the ECG measurements and especially the extracted HRVs. These differences are clearly shown in Table \ref{tab:datasets-feats}. The prevalence of the considered diseases is higher for both groups in MIMIC III. The sampled long-term HRVs of MIMIC III exhibit differences compared to Holter ECG data. Particularly, features, such as HR at rest, no longer show a  significant difference between no-HF and HF groups. Besides, MIMIC III lacks a future outcome for patients, which limits its use for classification only.
    
    \item \textbf{HF classification dataset:} {BIDMC \cite{baim1986survival} and NSRDB available in PhysioBank \cite{goldberger2000physiobank}, are commonly combined for HF classification studies \cite{acharya2019deep, kusuma2022ecg}. BIDMC compromises 15 subjects with severe \gls{HF} and 20-hour ECG recordings. NSRDB includes one ECG recording from 18 healthy individuals, respectively. The main limitation of these datasets is the reduced number of subjects.} 
    Therefore, it is a common practice to sample several ECG strips per subject. 
    Besides, the dataset lack information about the previous medical conditions of the patients.

    \item \textbf{Apple Watch data:} We collected several 30-second ECG recordings from 89 volunteers with Apple Watches and gathered their personal information and previous risk factors. Their future outcome is unknown. Thus, we are using their known risk factors and the atrial fibrillation labels given by the Apple Watch to corroborate that their predicted risks are higher than healthy individuals. The majority of the recordings, 331 samples from 84 subjects, exhibited sinus rhythm (SR). 17 recordings from 6 volunteers were tagged by the Apple Watch as high heart rate. And, 9 ECG strips from 3 subjects showed atrial fibrillation (AFib). The rest were considered inconclusive by the Apple Watch. Related to their known risk factors, 71 volunteers are healthy without any previous conditions, 11 subjects suffer only from hypertension (HTN), and 7 individuals have previous heart problems, such as arrhythmia (AFib and ectopic beats), ischemic heart disease, valvular heart disease, among others.  

\end{itemize}

The Holter ECG data is our primary dataset for training, validating, and testing our survival models. We split the data in a stratified manner, considering censoring information of the subjects, with 70\% allocated for training and validation, and 30\% reserved for testing. For hyperparameter tuning, we randomly searched through several possible configurations and selected the one with the highest concordance index (C-index) \cite{antolini2005time} evaluated in a 5-fold cross-validation of the training set. Then, we retrained them on the entire training set, and computed the final results on the test set. The rest of the datasets have been used exclusively for testing and no fine-tuning has been performed.

For survival settings, the model performance was evaluated using Antolini's C-index \cite{antolini2005time}, the areas under the ROC curves (AUC) within 5 years, the integrated brier score (iBS) \cite{graf1999assessment} and the average cumulative/dynamic AUC (c/d AUC) for all time horizons \cite{lambert2016summary}. When analyzing the discriminatory power using the classification datasets, we have used the AUC, the Average Precision (AP) of all probability thresholds for the Precision-Recall plot, the sensitivity, the specificity, and the geometric mean (G-mean). To test our models on the classification datasets, we are using the predicted probability of having HF within 5 years, which is the same time horizon used in our use case application.

\newcommand*{\tab}{\hspace*{0.2cm}}
\begin{sidewaystable}

\resizebox{\textwidth}{!}{
    \centering
\begin{tabular}{lcc|cc|cc|c}
\toprule
 &         \multicolumn{2}{c|}{\textbf{Holter ECG}}  &            \multicolumn{2}{c|}{\textbf{MIMIC-III}}  & \multicolumn{2}{c|}{\textbf{HF classification}} & \textbf{Watch}\\
 
 &            \textbf{No-HF} &               \textbf{HF} &            \textbf{No-HF} &               \textbf{HF} &           \textbf{No-HF} &               \textbf{HF} & \textbf{All} \\
\midrule

\textbf{Demographics}  &             &                &             &                &            &                & \\
\tab Sex                                   &             $50\%$ &             $57\%$ &             $59\%$ &             $50\%$ &            $26\%$ &             $76\%$ &            $72\%$ \\
\tab Age                                   &    $58.66\pm17.94$ &    $68.99\pm15.11$ &    $59.24\pm15.84$ &    $67.09\pm12.45$ &    $34.21\pm8.38$ &    $54.99\pm10.97$ &   $38.49\pm11.08$ \\

\textbf{Clinical history}   &             &                &             &                &            &                & \\
\tab Atrial fibrillation                   &              $8\%$ &             $26\%$ &             $21\%$ &             $54\%$ &             - &              - &             $2\%$ \\
\tab Chronic kidney disease                &             $10\%$ &             $28\%$ &             $13\%$ &             $52\%$ &             - &              - &             $1\%$ \\
\tab \begin{tabular}[c]{@{}c@{}} Chronic obstructive \\pulmonary disease\end{tabular} &             $11\%$ &             $23\%$ &              $9\%$ &             $34\%$ &             - &              - &             $0\%$ \\
\tab Diabetes mellitus                     &             $17\%$ &             $36\%$ &             $27\%$ &             $55\%$ &             - &              - &             $1\%$ \\
\tab Hyperlipidemia                        &             $22\%$ &             $35\%$ &             $37\%$ &             $60\%$ &             - &              - &             $1\%$ \\
\tab Hypertension                          &             $37\%$ &             $62\%$ &             $50\%$ &             $61\%$ &             - &              - &             $9\%$ \\
\tab Ischemic heart disease                &             $16\%$ &             $40\%$ &             $24\%$ &             $63\%$ &             - &              - &             $2\%$ \\
\tab Myocardial infarction                  &              $2\%$ &             $12\%$ &             $12\%$ &             $44\%$ &             - &              - &             $0\%$ \\
\tab Stroke                                &              $9\%$ &             $16\%$ &             $20\%$ &             $14\%$ &             - &              - &             $0\%$ \\
\tab Valvular heart disease                &              $6\%$ &             $18\%$ &              $2\%$ &             $15\%$ &             - &              - &             $1\%$ \\

\textbf{ECG Rhythm}  &             &                &             &                &            &                \\
\tab Mean HR (BPM)                         &    $76.87\pm20.31$ &    $80.91\pm23.94$ &    $89.66\pm20.89$ &    $88.91\pm19.32$ &   $75.43\pm13.89$ &    $90.85\pm16.95$ &   $77.46\pm15.86$ \\
\tab SD1/SD2                               &      $0.74\pm1.03$ &      $1.03\pm1.16$ &      $0.92\pm3.34$ &      $1.08\pm0.82$ &     $0.48\pm0.23$ &      $0.89\pm0.53$ &     $0.67\pm0.44$ \\
\tab SDNN (ms.)                            &    $60.01\pm84.63$ &    $90.39\pm95.72$ &    $48.46\pm61.21$ &    $65.55\pm65.58$ &   $51.02\pm33.07$ &    $40.17\pm54.51$ &   $96.72\pm109.4$ \\

\textbf{ECG Shape}  &             &                &             &                &            &                \\
\tab P timing (ms.)                        &   $145.85\pm44.36$ &   $158.89\pm63.99$ &   $136.41\pm43.42$ &   $144.83\pm54.93$ &  $141.99\pm36.06$ &   $158.41\pm36.13$ &  $133.72\pm51.89$ \\
\tab Q timing (ms.)                        &    $36.86\pm13.67$ &    $39.72\pm15.26$ &     $39.54\pm15.1$ &    $41.31\pm18.62$ &   $46.51\pm22.77$ &    $33.46\pm18.48$ &    $36.7\pm11.52$ \\
\tab R timing (ms.)                        &  $434.62\pm101.86$ &  $425.65\pm111.52$ &   $367.46\pm89.56$ &    $368.76\pm71.0$ &   $439.0\pm85.47$ &   $350.09\pm71.87$ &  $426.01\pm68.09$ \\
\tab S timing (ms.)                        &    $28.39\pm14.58$ &    $31.77\pm15.46$ &    $36.88\pm16.81$ &    $42.67\pm19.42$ &    $26.49\pm4.77$ &    $37.09\pm10.39$ &    $25.11\pm5.47$ \\
\tab T timing (ms.)                        &   $266.66\pm54.98$ &   $269.89\pm71.24$ &    $244.3\pm54.34$ &   $249.38\pm67.75$ &   $264.02\pm41.8$ &   $239.46\pm51.44$ &  $234.09\pm29.97$ \\
\tab P amplitude (mV.)                     &      $0.02\pm0.06$ &      $0.02\pm0.05$ &       $0.1\pm0.72$ &      $0.03\pm0.08$ &     $0.08\pm0.05$ &      $0.12\pm0.18$ &     $0.04\pm0.04$ \\
\tab Q amplitude (mV.)                     &      $-0.03\pm0.1$ &     $-0.04\pm0.18$ &     $-0.07\pm0.38$ &     $-0.04\pm0.11$ &    $-0.06\pm0.07$ &     $-0.07\pm0.11$ &    $-0.05\pm0.05$ \\
\tab R amplitude (mV.)                     &      $0.59\pm0.35$ &      $0.61\pm0.47$ &      $0.71\pm4.25$ &      $0.47\pm0.32$ &     $0.46\pm0.31$ &      $0.45\pm0.86$ &     $0.49\pm0.33$ \\
\tab S amplitude (mV.)                     &      $-0.2\pm0.21$ &     $-0.27\pm0.29$ &      $-0.3\pm2.87$ &      $-0.2\pm0.21$ &    $-0.83\pm0.38$ &      $-1.6\pm1.17$ &     $-0.14\pm0.1$ \\
\tab T amplitude (mV.)                     &       $0.1\pm0.13$ &      $0.04\pm0.15$ &      $0.15\pm1.24$ &      $0.02\pm0.15$ &      $0.25\pm0.3$ &      $0.31\pm0.44$ &     $0.19\pm0.08$ \\

\textbf{Sampled long-term HRV}  &             &                &             &                &            &                \\
\tab HR at rest (BPM)                      &    $66.24\pm12.26$ &     $71.9\pm16.01$ &    $78.06\pm15.39$ &    $78.43\pm12.97$ &   $77.03\pm11.96$ &    $92.87\pm16.42$ &       - \\
\tab Active HR (BPM)                      &    $84.52\pm15.13$ &    $85.14\pm19.18$ &    $91.58\pm17.18$ &    $91.22\pm15.68$ &   $91.86\pm15.07$ &    $100.64\pm15.9$ &       - \\
\tab SDNN (ms.)                            &   $145.21\pm68.28$ &     $150.8\pm83.8$ &  $143.42\pm306.51$ &   $165.96\pm100.5$ &   $98.09\pm33.47$ &    $86.71\pm158.2$ &       - \\
\tab RMSSD                                 &  $102.09\pm104.16$ &   $161.03\pm117.1$ &  $157.86\pm435.21$ &  $203.84\pm143.67$ &   $54.42\pm30.09$ &  $106.51\pm207.69$ &       - \\
\tab PNN50 (\%)                             &      $18.3\pm22.4$ &    $34.58\pm29.98$ &    $17.97\pm19.94$ &    $30.88\pm25.83$ &    $13.01\pm9.86$ &    $12.61\pm18.02$ &       - \\
\tab SD1/SD2                               &      $4.87\pm3.55$ &      $2.41\pm2.31$ &      $2.16\pm1.78$ &      $1.46\pm0.61$ &     $4.08\pm1.54$ &      $1.78\pm1.19$ &       - \\
\tab Ellipse's area                        &    $5e04\pm9e04$ &    $8e04\pm1e05$ &    $4e05\pm1e07$ &    $1e05\pm3e05$ &   $2e04\pm1e04$ &    $1e05\pm4e05$ &       - \\
\tab Sample Entropy                        &       $0.53\pm0.4$ &      $0.76\pm0.64$ &      $0.55\pm0.41$ &      $0.65\pm0.53$ &     $0.83\pm0.29$ &      $0.65\pm0.35$ &       - \\
\bottomrule
\end{tabular}
}
    \caption{Statistics (\% or mean$\pm$std) of the extracted features of each dataset stratified by HF observation.} 
    \label{tab:datasets-feats}
\end{sidewaystable}

\subsection{Model performance in survival analysis settings}
\label{subs:exp-performance}

Table \ref{tab:res-models} shows the performance of 16 different models on the Holter ECG test set. We have grouped them into four categories. The traditional category includes a baseline model trained only with the demographic information and comorbidities of the subjects and the classic survival models, \gls{CPH} and \gls{AFT} \cite{kleinbaum2010survival}. The features-based models are MLP DeepHit \cite{lee2018deephit}, \gls{RSF} \cite{10.1214/09-SS047}, \gls{EST} \cite{10.1214/09-SS047}, and our XGBoost AFT \cite{gonzalez2020practical}. As for signal-based models, we consider four deep learning architecture for time series: ResCNN \cite{zou2019integration}, XceptionTime \cite{rahimian2019xceptiontime}, InceptionTime \cite{ismail2020inceptiontime}, and our ResNet \cite{he2016deep}. The multi-modal approaches are based on ResNet or InceptionTime, the two best-performing ECG feature extractors, and GRU or Transformer (TFM) as HRV series model. Note that all machine learning approaches use our ECG features described in Section \ref{subsec:feat-model}, while the deep learning models have been adapted according to our survival analysis methodology in Section \ref{subsec:raw-model}. 

{Besides, we extended all these models to learn from HRV information as described in Section \ref{subsec:hrvs}. Traditional, features-based, and signal-based models include an additional HRV features vector extracted from the concatenation of the different beat-to-beat measurements. Deep learning multi-modal models incorporate a sequence model, such as GRU or TFM, to learn from HRV time series extracted from each beat-to-beat measurement independently. Their performance results with our approximated long-term HRV are in Table \ref{tab:res-models-app}.}

As shown in Table \ref{tab:res-models}, all models with 30s ECG information significantly outperform the baseline model only with the subject's clinical information. More importantly, our sampled long-time HRVs consistently boost model performance, as evidenced by the improved results of all HRV-based models compared to their counterparts in Table \ref{tab:res-models-app}. 

Within each category, the best-performing models are CPH, XGBoost AFT, ResNet, and our TFM-ResNet. The traditional CPH method is clearly the worst comparing across categories, followed by XGBoost AFT and then ResNet. Finally, our TFM-ResNet model outperforms the rest, including the HRV-boosted models, in all metrics except for the iBS of RSF.

\begin{figure}
\centering 
\includegraphics[width=\textwidth]{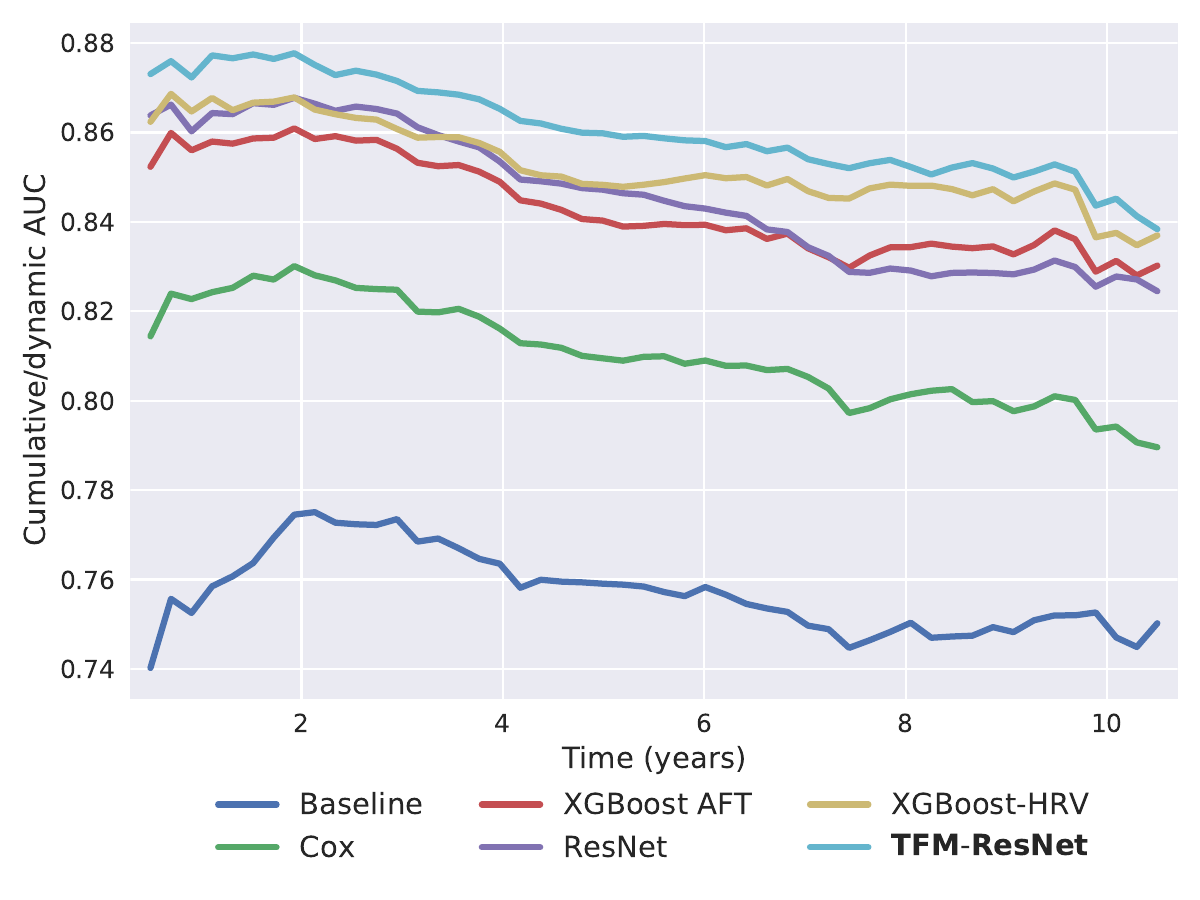} 
\caption{\color{black}Cumulative/dynamic AUC of the best survival models of each category at each time horizon corrected for right-censored time-to-event data.} 
\label{fig:cdAUC}
\end{figure}

Figure \ref{fig:cdAUC} shows the cumulative/dynamic AUC over time, which exhibits the performance variation over different time horizons. The peak performance of all models is for predictions within two years when most of our HF cases occur (Avg. time-to-event: 1.9). Then, the model performance decreases for greater time horizons. Interestingly, models trained with the sampled long-term HRVs experience a less abrupt decrease (TFM-ResNet) or even a reincrease in performance (XGBoost-HRV). Figure \ref{fig:cdAUC} also leads to similar conclusions as the previous table: our TFM-ResNet is the best model with an AUC c/d curve over the rest, and sampled long-term HRVs enhance the model performance of the 30s ECG models.

\begin{table}

\resizebox{\textwidth}{!}{
\centering 
\begin{tabular}{llcccc}
\toprule
\textbf{Category} & \textbf{Model} &  \textbf{C-index} &    \textbf{AUC} ($y\le5$) &     \textbf{iBS} &  \textbf{c/d AUC} \\
\midrule
\multirow{3}{*}{Traditional}&Baseline         &   0.7476 &  0.7608 &  0.1073 &   0.7538 \\
&AFT              &   0.8009 &  0.8128 &  0.0664 &   0.8168 \\
&CPH              &   0.8010 &  0.8129 &  0.0669 &   0.8166 \\
\midrule
\multirow{4}{*}{Features-based}&MLP DeepHit      &   0.8025 &  0.8171 &  0.0691 &   0.8220 \\
&RSF               &   0.8217 &  0.8369 &  \textbf{0.0632} &   0.8396 \\
&EST               &   0.8052 &  0.8235 &  0.0644 &   0.8237 \\
&\textbf{XGBoost AFT}         &   0.8312 &  0.8440 &  0.0921 &   0.8508 \\
\midrule
\multirow{4}{*}{Signal-based}&ResCNN           &   0.8163 &  0.8307 &  0.0771 &   0.8318 \\
&XceptionTime     &   0.8302 &  0.8429 &  0.0821 &   0.8479 \\
&InceptionTime    &   0.8366 &  0.8490 &  0.0863 &   0.8556 \\
&\textbf{ResNet}           &   0.8403 &  0.8545 &  0.0823 &   0.8578 \\
\midrule
\multirow{4}{*}{Multi-modal} & \color{black}XGBoost-HRV  & \color{black}0.8400 &  \color{black}0.8517 &  \color{black}0.0901 &  \color{black}0.8598 \\
&GRU-InceptionTime &   0.8441 &  0.8586 &  0.0734 &   0.8674 \\
&GRU-ResNet       &   0.8469 &  0.8592 &  0.0767 &   0.8642 \\
&TFM-InceptionTime &   0.8496 &  0.8649 &  0.0777 &   0.8690 \\
&\textbf{TFM-ResNet}       &   \textbf{0.8537} &  \textbf{0.8688} &  0.0744 &   \textbf{0.8718} \\
\bottomrule
\end{tabular}
}
\caption{Performance of the survival models on the Holter ECG test set.}
\label{tab:res-models}
\end{table}

\subsection{Approximating long-term HRV: Sample size and frequency}
\label{subs:exp-hrv}

This experiment shows the effect of including additional beat-to-beat information when computing our approximate long-term HRVs. To do so, we have varied both the size of the beat-to-beat samples and their frequency during the day. Concerning the sample size, we have examined 1, 5, 10, 30 minutes, and 1 hour (whole day), fixing the number of samples to 24 (1 per hour). As for the number of sampled beat-to-beat measurements, we have considered 3, 6, 12, and 24, keeping their size equal to one minute. 

\begin{figure}
	\begin{center}
    \subfloat[\label{fig:corr-hrv-size}]{\includegraphics[width=0.5\textwidth]{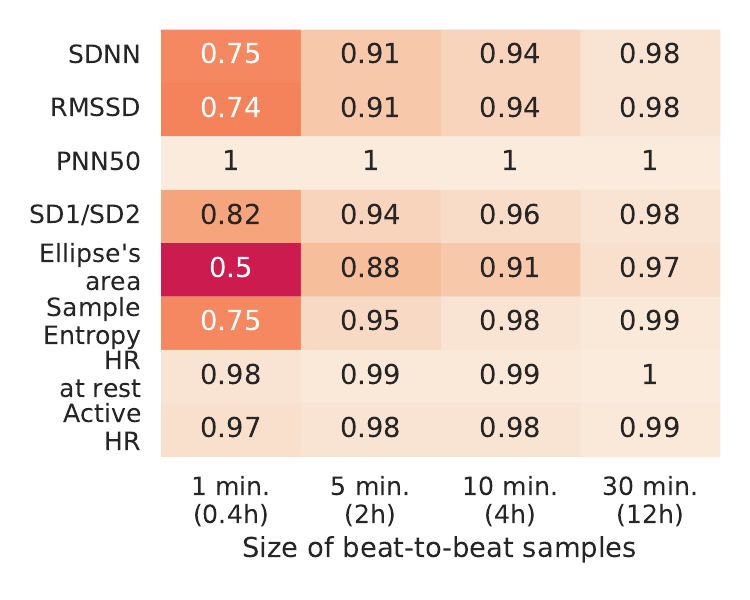}}
	\subfloat[\label{fig:corr-hrv-num}]{\includegraphics[width=0.5\textwidth]{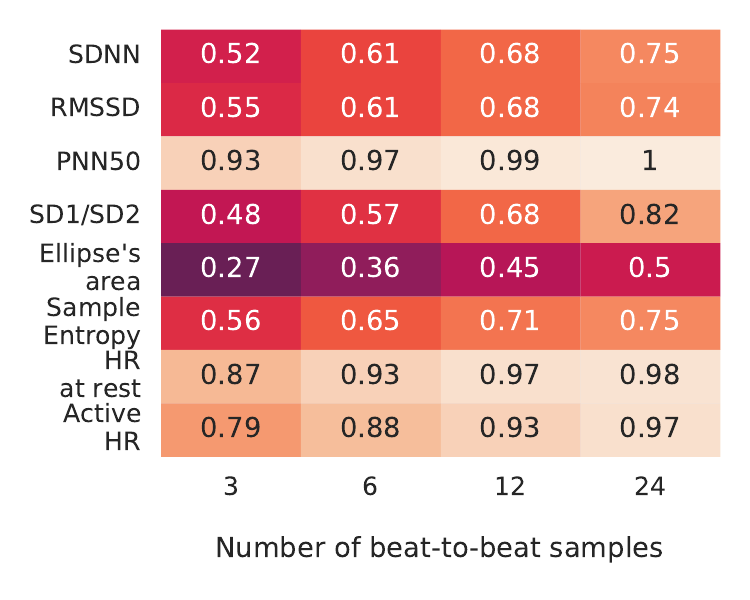}}
    \caption{Correlation of the HRV features extracted from different sample sizes (a) and numbers (b) with those extracted from 24 hours}
	\label{fig:corr-hrv}		
	\end{center}
\end{figure}

Figure \ref{fig:corr-hrv} shows the correlation of the HRV features extracted from different sample sizes (\ref{fig:corr-hrv-size}) and numbers (\ref{fig:corr-hrv-num}) compared to those extracted from the complete 24 hours. As expected, the correlation increases with a larger sample size and more measurements. However, the correlation is already considerably high with 24 samples of 1-minute size and above 0.9 with 5-minute in Figure \ref{fig:corr-hrv-size}. With limited sample numbers such as 3 or 6 in Figure \ref{fig:corr-hrv-num}, the correlations notably decrease except for PNN50, Active HR, and HR at rest. This is already an indication that we could approximate long-term HRVs from ultra-short measurements ($\le$ 5 min.), provided we take about 12 to 24 random samples per day.

\begin{figure}
	\begin{center}
    \subfloat[\label{fig:per-hrv-size}]{\includegraphics[width=0.5\textwidth]{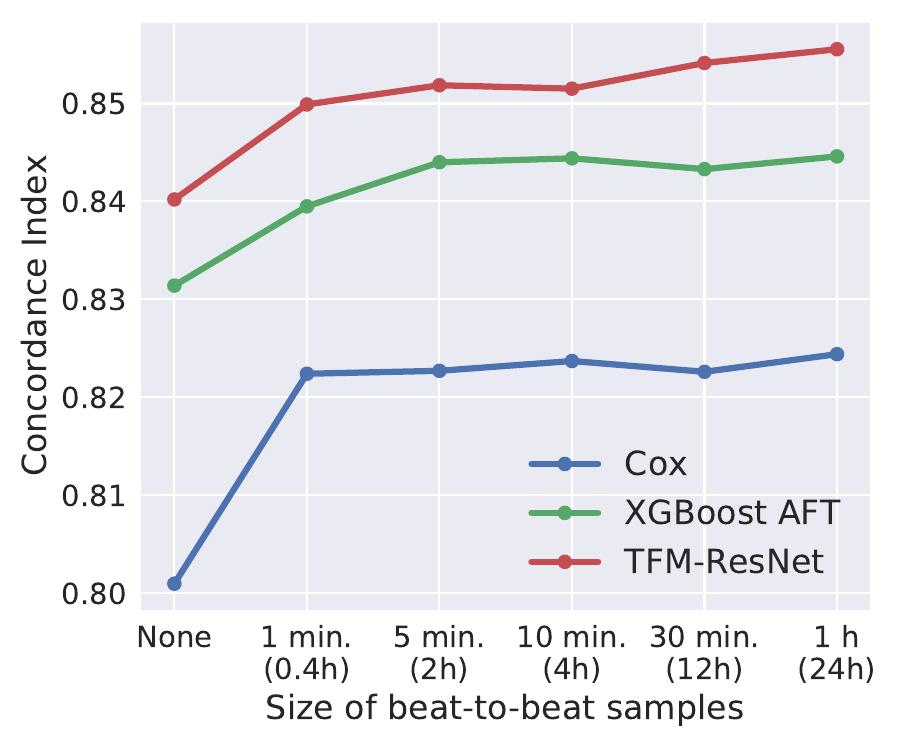}}
	\subfloat[\label{fig:per-hrv-num}]{\includegraphics[width=0.5\textwidth]{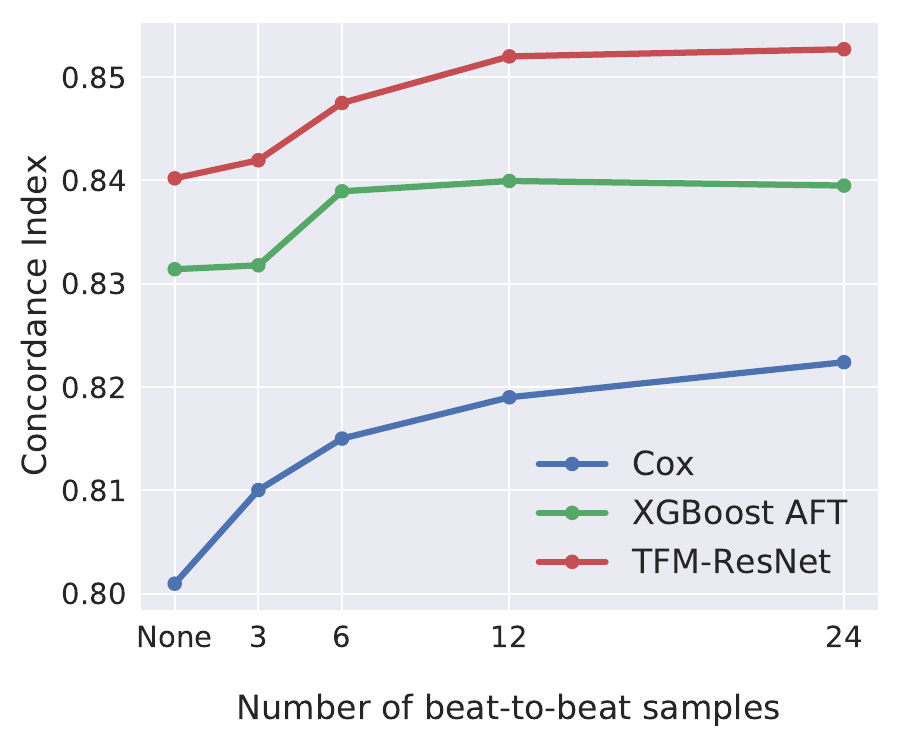}}
    \caption{Model performance with different sizes (a) and numbers (b) of beat-to-beat measurements when computing our sampled long-term HRVs.}
	\label{fig:per-hrv}		
	\end{center}
\end{figure}

{Figure \ref{fig:per-hrv} exhibits the C-index of CPH, our XGBoost AFT, and our TFM-ResNet with different sizes (\ref{fig:per-hrv-size}) and numbers (\ref{fig:per-hrv-num}) of beat-to-beat measurements.} As aforementioned, our approximate HRVs always help improve the model performance regardless of the limited information. Concerning only sample size, one-minute samples are sufficient to achieve a performance boost comparable to that of complete 24-hour HRVs. However, the number of considered measurements produces more gradual changes in model performance. As shown in Figure \ref{fig:per-hrv-num}, the performance improvement with only three samples is limited, but it increases with the number of measurements and plateaus with 12 or more. 

This outcome supports our idea of approximating long-term HRVs to boost the model performance with a greater understanding of heart activity during the day. Besides, using limited information without performance loss enables some potential advantages in its implementation. For example, reducing data processing would extend the battery life of the device. HRVs can be extracted from PPG sensors of smartwatches facilitating their implementation and use. As a 24-hour measurement is not required, the wearable could filter out poor-quality measurements caused by motion artifacts or other reasons. Additionally, the user would not need to wear it throughout the day. Finally, some current wearables, such as the Apple Watch, only measure a few beat-to-beat measurements per day. Thus, our strategy could be seamlessly implemented on such devices.

\subsection{Transferability to open ECG datasets}
\label{subs:exp-opendata}

{With this experiment, we aim to evaluate our models with other external ECG datasets. Since HF survival data are costly and difficult to obtain, we evaluated the discriminative capabilities of our models in the HF classification task with MIMIC III, and HF classification datasets (BIDMC and NSRDB). That is, whether our models estimate significantly lower risks for healthy than for diseased individuals. Thus, we have used the probability of having HF within five years predicted by our models as the classification discriminator. Note that the models have not been retrained or fine-tuned for these datasets.}

\begin{figure}
	\begin{center}
    \subfloat[\label{fig:roc-hfclass}]{\includegraphics[width=0.5\textwidth]{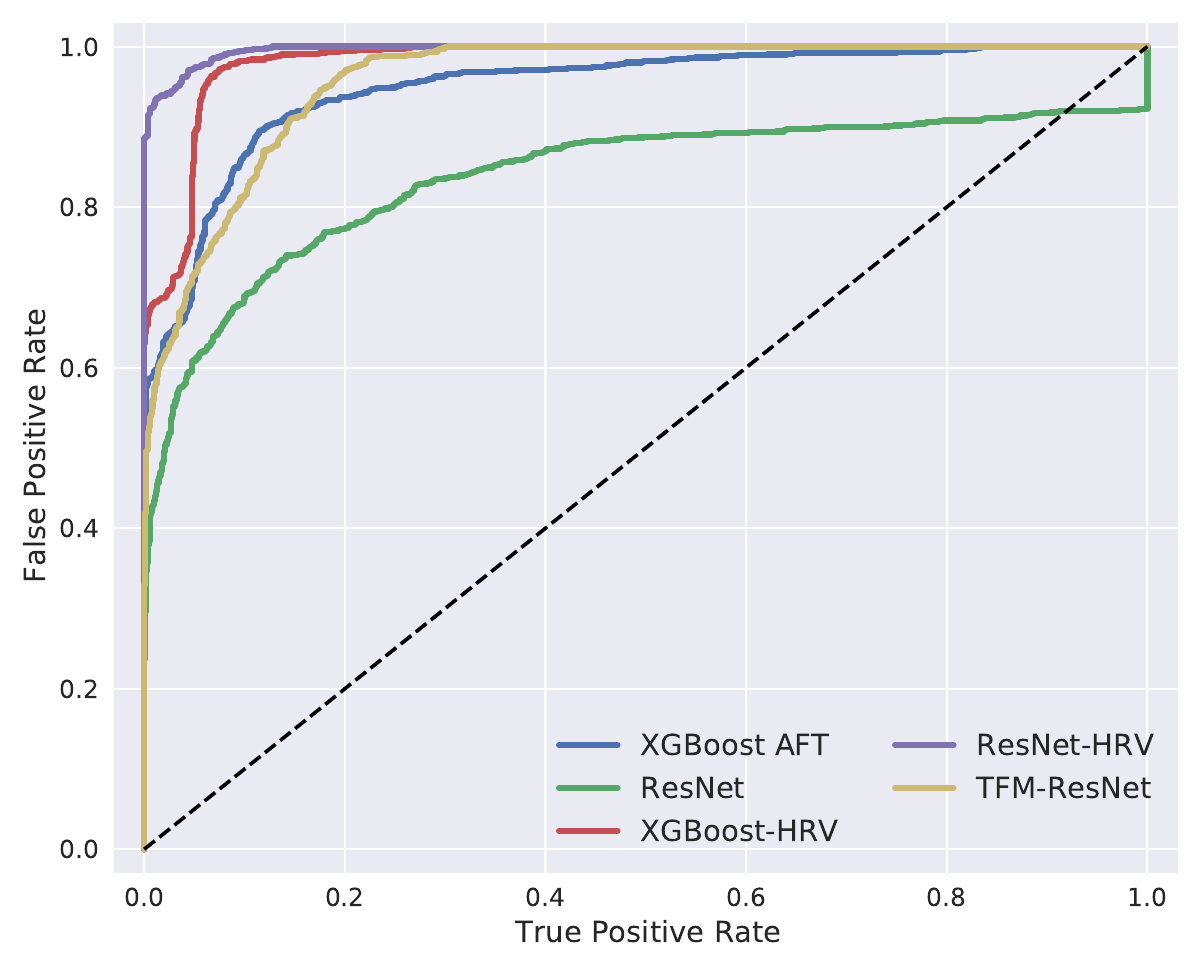}}
	\subfloat[\label{fig:ap-hfclass}]{\includegraphics[width=0.5\textwidth]{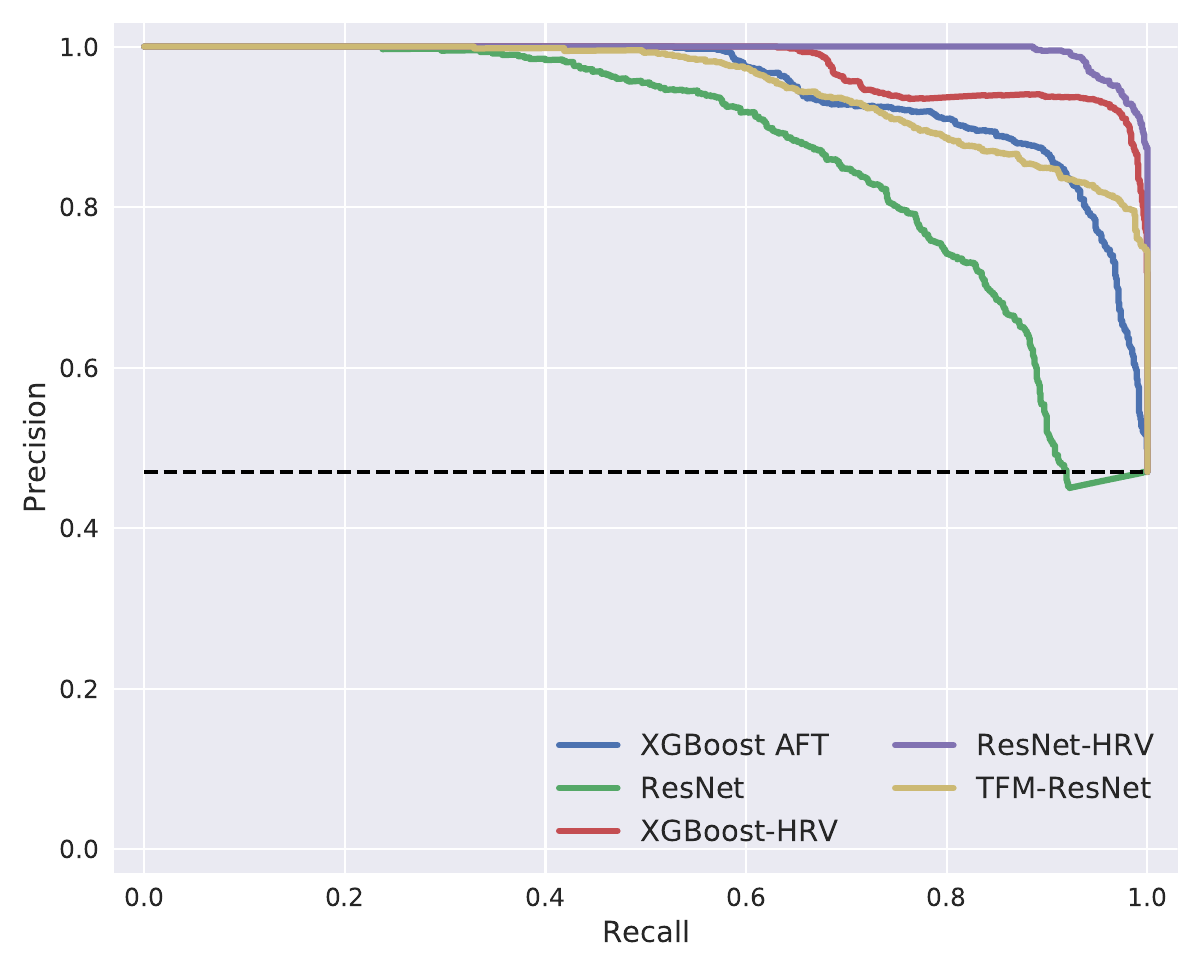}}
    \caption{ROC (a) and Precision-Recall (b) curves of our models for HF classification data. The dashed line represents chance performance.}
	\label{fig:classHF}		
	\end{center}
\end{figure}

Table \ref{tab:res-models-class} gathers the results of our models on MIMIC-III and the HF classification datasets. In general, our models exhibit good discrimination discerning between subjects with and without HF on these datasets. Particularly, the results achieved on the HF classification dataset are comparable to those of the state-of-the-art in this task \cite{acharya2019deep, kusuma2022ecg}. The models trained with our sampled long-term HRVs outperform those trained only with 30s ECG. This difference in performance is less pronounced in MIMIC-III because, as discussed above and shown in Table \ref{tab:datasets-feats}, the situation of ICU patients substantially affects the extracted HRVs. On the other hand, the difference is evident in the HF classification set, where the ROC and Precision-Recall curves of the models with HRVs in Figure \ref{fig:classHF} are above the rest.

\begin{table}
\centering

\begin{tabular}{lccccc}
\toprule
(MIMIC-III) &     AUC &      AP &  Sensitivity &  Specificity &  G-mean \\
\midrule
XGBoost AFT        &  0.8219 &  0.3548 &       0.7583 &       0.7482 &  0.7533 \\
ResNet         &  0.8196 &  0.3650 &       0.7915 &       0.6970 &  0.7427 \\
\midrule
XGBoost-HRV    &  0.8072 &  0.3329 &       \textbf{0.8057} &       0.6744 &  0.7371 \\
ResNet-HRV     &  \textbf{0.8343} &  \textbf{0.3939} &       0.7014 &       \textbf{0.8186} &  0.7577 \\
TFM-ResNet     &  0.8152 &  0.3599 &       0.7333 &       0.7838 &  \textbf{0.7581} \\
\midrule
\midrule
(HF classification) &     AUC &      AP &  Sensitivity &  Specificity &  G-mean \\
\midrule
XGBoost AFT    &  0.9437 &  0.9439 &       0.8980 &       0.8596 &  0.8786 \\
ResNet      &  0.8373 &  0.8729 &       0.7397 &       0.8582 &  0.7968 \\
\midrule
XGBoost-HRV &  0.9818 &  0.9779 &       \textbf{0.9802} &       0.9169 &  0.9481 \\
ResNet-HRV  &  \textbf{0.9963} &  \textbf{0.9960} &       0.9703 &       \textbf{0.9554} &  \textbf{0.9628} \\
TFM-ResNet  &  0.9584 &  0.9522 &       0.9444 &       0.8258 &  0.8831 \\
\bottomrule
\end{tabular}
\caption{Performance of our HF models with and without our sampled long-term HRV on open datasets.}
\label{tab:res-models-class}
\end{table}

\subsection{Transferability to wearable devices: Apple Watch}
\label{subs:exp-watch}

{We are committed to implementing our proposals into existing wearables, such as the Apple Watch, to help uncover potential underlying risks of the users. Therefore, we aim with this experiment to verify that our models perform without significant discrepancies or major errors when applied to Apple Watch's data. Thus, we recorded several 30-second ECGs from 89 volunteers with a few Apple Watches. Since their future outcome is unknown, we compare the predicted survival curves of healthy subjects and those with atrial fibrillation or known cardiac risk factors. Although we could not include HRV data in this experiment due to limitations in our collection process, we expect that HRVs computed from Apple Watch's beat-to-beat measurements would only improve the results, as shown in previous experiments.}

\begin{figure}
	\begin{center}

     \subfloat[\label{fig:tag-xgb}]{\includegraphics[width=0.5\textwidth]{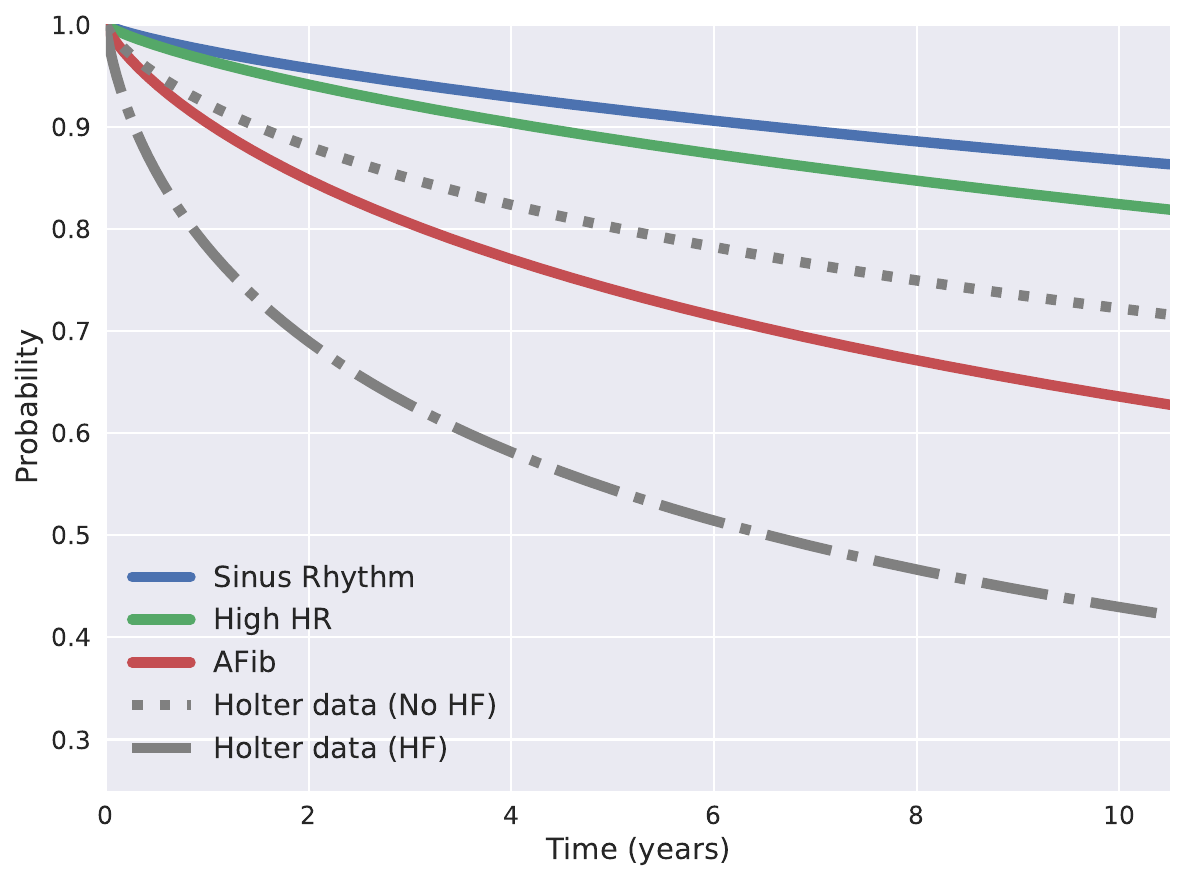}}
    \subfloat[\label{fig:tag-rsnt}]{\includegraphics[width=0.5\textwidth]{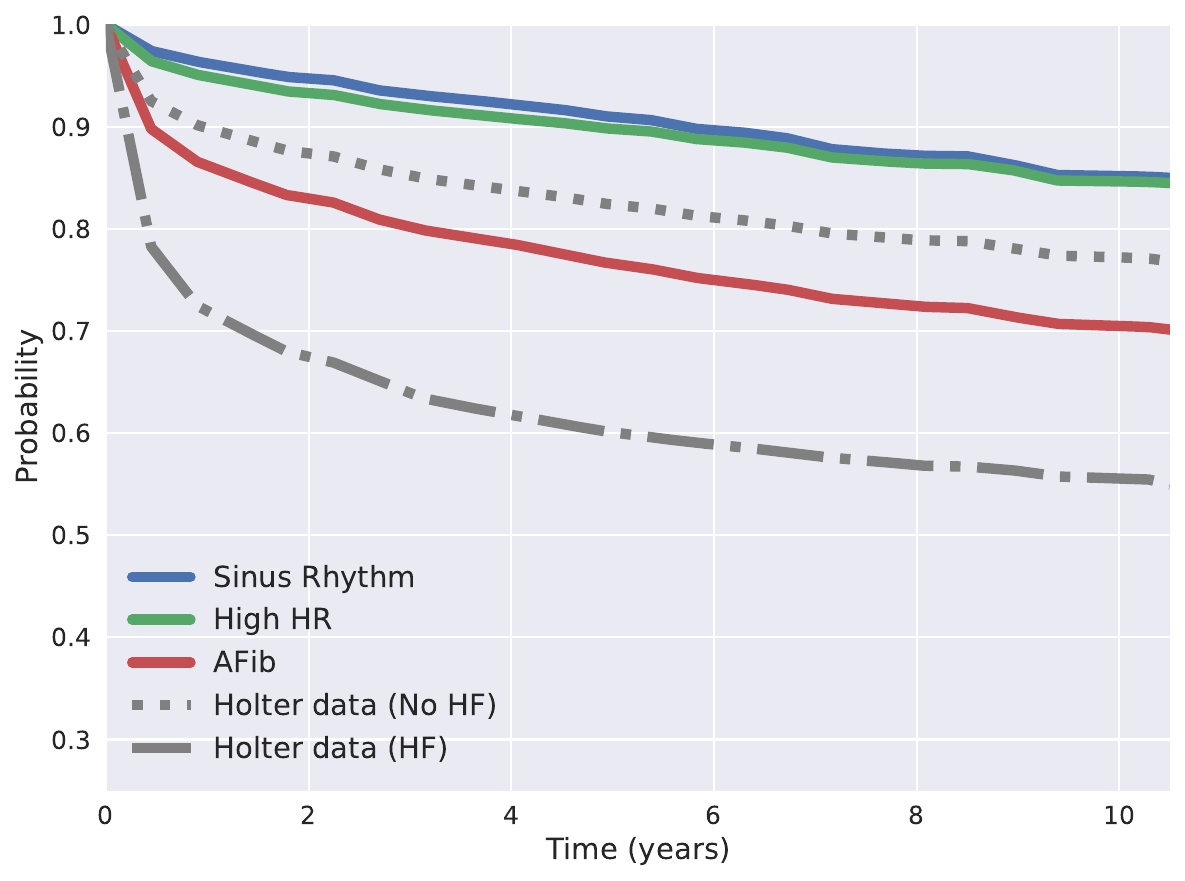}}
 
    \subfloat[\label{fig:risk-xgb}]{\includegraphics[width=0.5\textwidth]{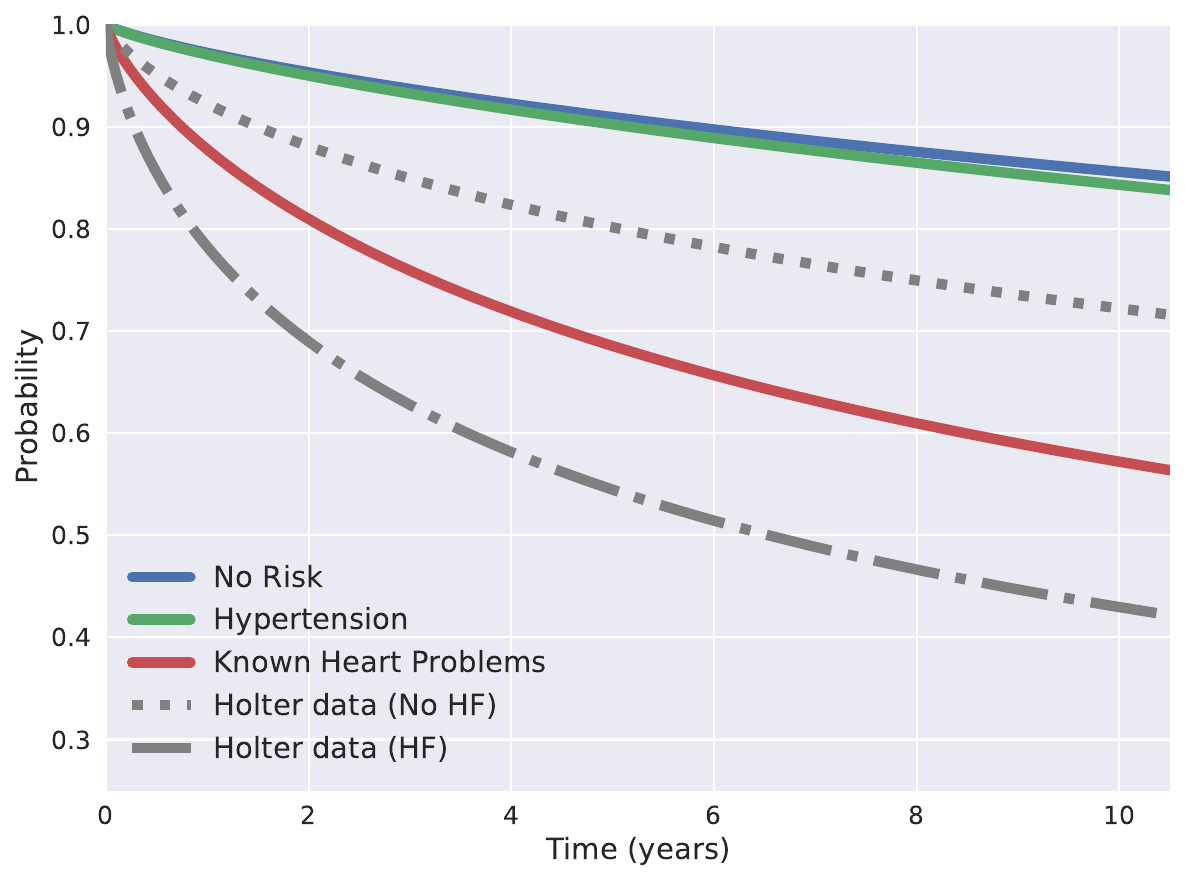}}
	\subfloat[\label{fig:risk-rsnt}]{\includegraphics[width=0.5\textwidth]{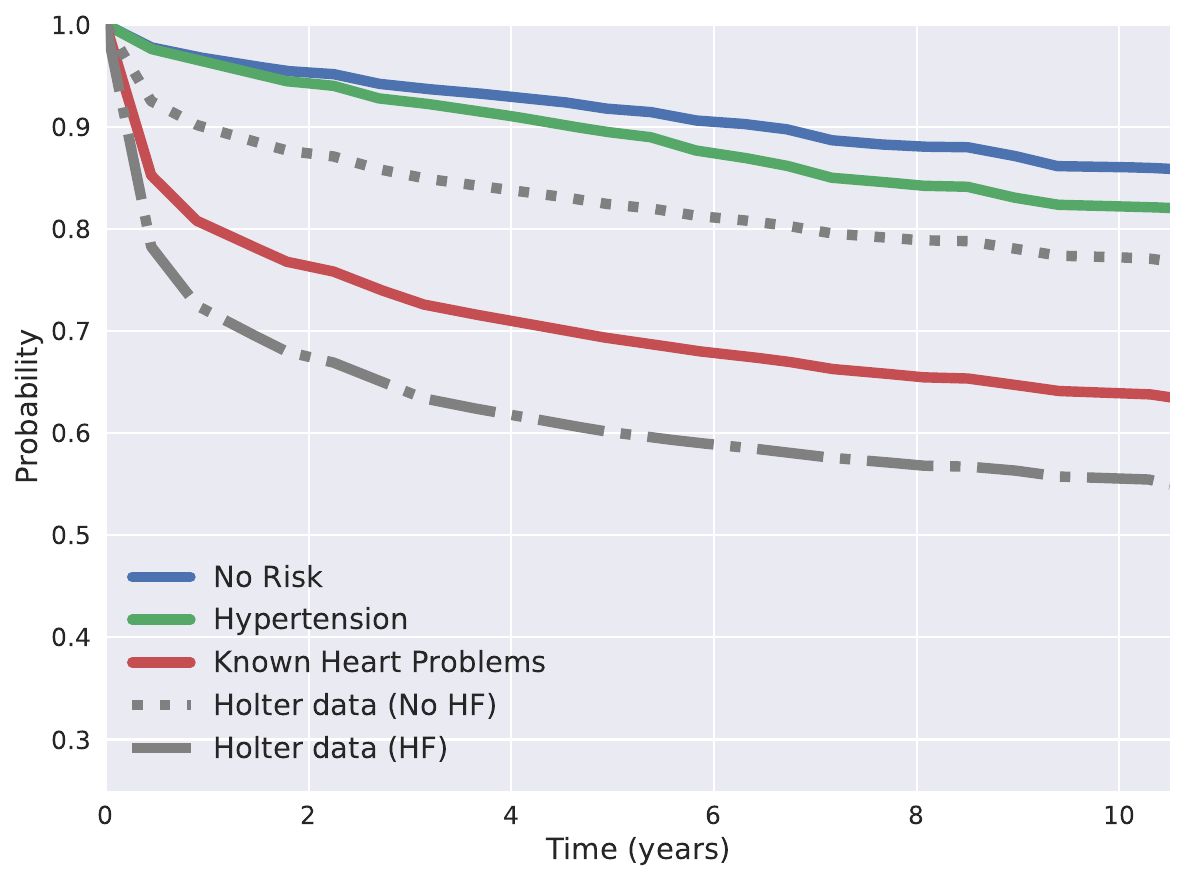}}
    
    \caption{Survival curves predicted by our XGBoost AFT (left) and our ResNet (right) aggregated by Apple Watch's tags (a \& b) and the known risk factors (c \& d). For reference, the grey lines are the average survival curves for the individuals with and without HF of the Holter ECG test set.}
	\label{fig:surv-AW}		
	\end{center}
\end{figure}

Figure \ref{fig:surv-AW} shows the average predictions of our models for the ECG labels given by the Apple Watch (a \& b) and the known medical conditions of the subjects (c \& d). These survival curves are predicted by our XGBoost AFT and ResNet models without our sampled long-term HRVs.

First, Figure \ref{fig:surv-AW} reveals an overall agreement between the models in their predictions of the different groups. Although ResNet makes more conservative predictions. Regarding the labels given by Apple Watch (Figures \ref{fig:tag-xgb} \& \ref{fig:tag-rsnt}), our models give slightly lower probabilities for those ECG strips with high heart rate compared to sinus rhythm ECGs. Furthermore, atrial fibrillation cases scored significantly lower, dropping below the mean predicted probability for subjects without HF. 

Besides, we are comparing the predicted curves across different risk categories for the subjects, which are grouped in no risk, hypertensive individuals, and subjects with previous cardiovascular problems. As shown in Figure \ref{fig:risk-xgb} \& \ref{fig:risk-rsnt}, hypertensive subjects have similar or slightly lower curves on average compared to the group without known risk factors. However, the predictions for people with known heart problems are significantly lower. This behavior is reasonable since this group comprises persons with illnesses such as ischemic heart and valvular heart diseases. 

Figure \ref{fig:examples} shows examples of the recordings included in this experiment and their probabilities of remaining healthy within 5 years predicted by our models, i.e., 100\% - the probability of HF. They are presented from lower to higher predicted probabilities. The subject with the lowest average probability in Figure \ref{fig:example0} is an 86-year-old female previously diagnosed with severe septal hypertrophy in the left ventricle and moderate mitral insufficiency. The individual also suffers from hypertension and ischemic heart disease. Its ECG recording shows a low heart rate paced by a pacemaker. 

Figure \ref{fig:example1} exhibits an atrial fibrillation recording of a male with 68 years old and diagnosed with atrial fibrillation, hypertension, ischemic, and valvular heart diseases. The subject recently underwent surgery due to excessive aortic enlargement and stenosis of his bicuspid aortic valve. This ECG was performed one month after the replacement of his valve and part of the aorta. 

Next, the ECG of a \textcolor{black}{36-year-old} male with diabetes and atrial fibrillation is presented in Figure \ref{fig:example2}. It exhibits an inverted R wave and an irregular rhythm with a longer or later beat every 4 to 6 beats, which our models penalize with low probabilities. This individual required multiple heart surgeries to correct a congenital heart defect and currently wears a pacemaker. At present, the subject is stable and is preparing for amateur triathlon competitions.

Figure \ref{fig:example3} shows an atrial fibrillation recording of a male with 68 years old previously diagnosed with atrial fibrillation and hypertension. Without additional health problems, our models predict higher probabilities compared to the previous cases. The ECG in Figure \ref{fig:example4} illustrates an arrhythmia with trigeminy (ectopic beats) in a 37-year-old man. His predictions are within 80-89\%. Finally, the young male with sinus rhythm of Figure \ref{fig:example5} receives considerably high predicted probabilities. As the above results and examples show, the probabilities predicted by our models seem to be correlated with the health states and medical conditions of the subjects.

\newcommand*{\exapsize}{\textwidth}
\begin{figure}

	\begin{center}
     \subfloat[86 y/o Female, HTN, IHD. Watch: Low Heart Rate, XGBoost: 49\%, ResNet: 62 \% \label{fig:example0}]{\includegraphics[width=\exapsize]{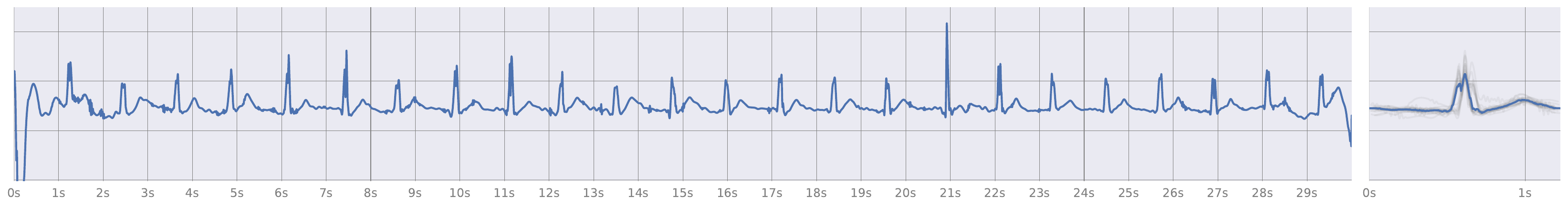}}
     
     \subfloat[68 y/o Male, AF, HTN, IHD, VHD. Watch: Atrial Fibrillation, XGBoost: 56\%, ResNet: 56\%\label{fig:example1}]{\includegraphics[width=\exapsize]{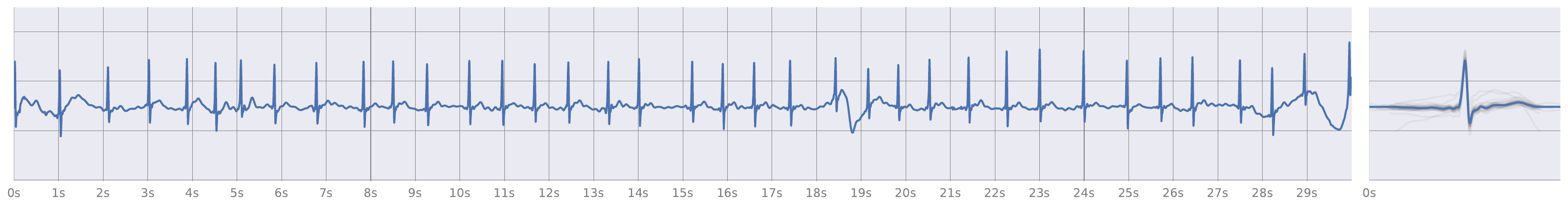}}
     
     \subfloat[36 y/o Male, AF, DM. Watch: Sinus Rhythm, XGBoost: 57\%, ResNet: 62\%\label{fig:example2}]{\includegraphics[width=\exapsize]{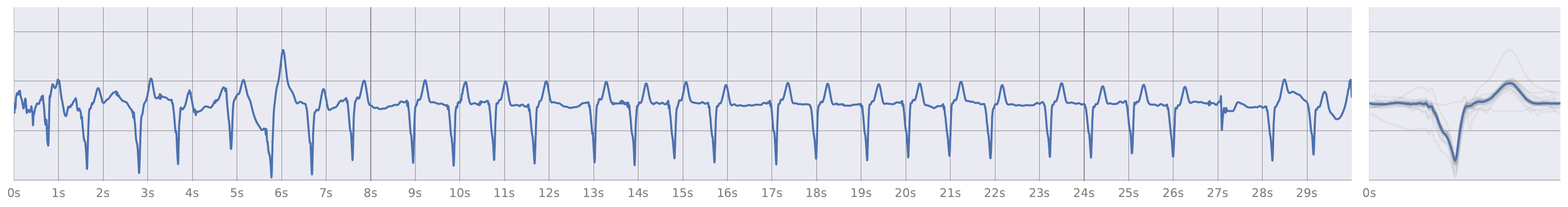}}
     
     \subfloat[61 y/o Male, AF, HTN. Watch: Atrial Fibrillation, XGBoost: 72\%, ResNet: 79\%\label{fig:example3}]{\includegraphics[width=\exapsize]{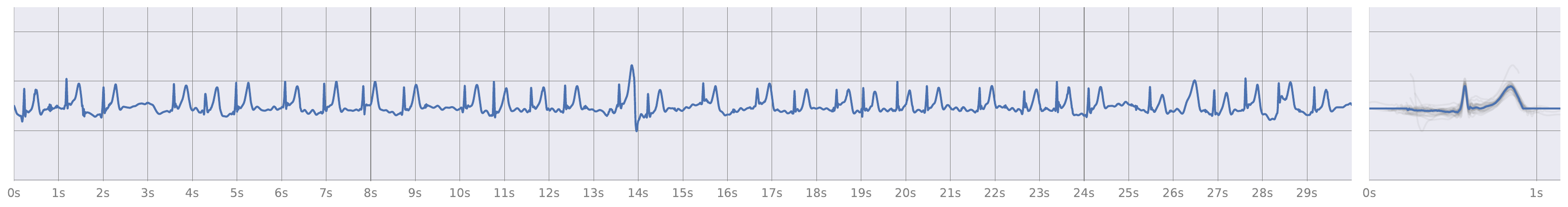}}
     
     \subfloat[37 y/o Male, Watch: Inconclusive, XGBoost: 82\%, ResNet: 88\%\label{fig:example4}]{\includegraphics[width=\exapsize]{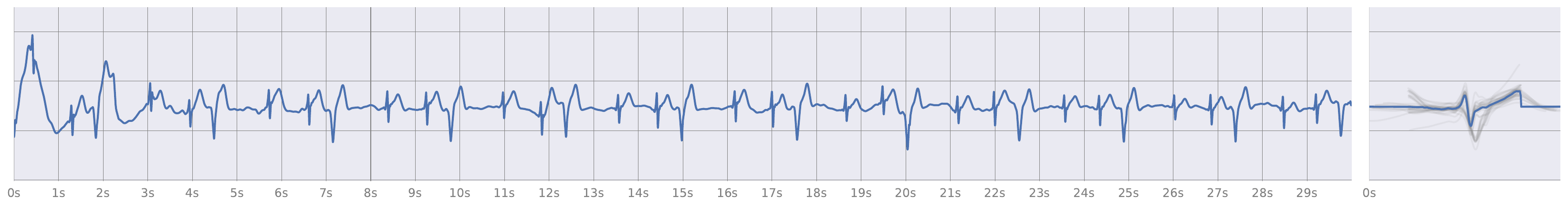}}
     
     \subfloat[30 y/o Male, Watch: Sinus Rhythm, XGBoost: 95\%, ResNet: 94\% \label{fig:example5}]{\includegraphics[width=\exapsize]{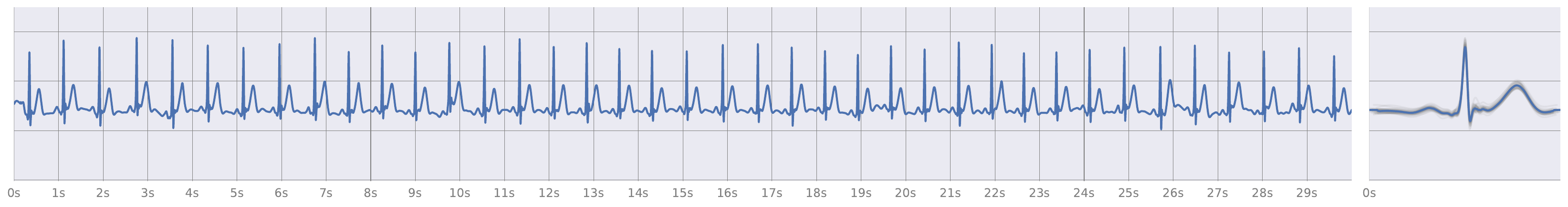}}
    \caption{Some individuals, their ECG recordings, and cycle templates included in our Apple Watch study. The captions include the sex, age (year-old -- y/o), and previous medical conditions of the subjects, the label given by the Apple Watch, and the probabilities of remaining free from HF with 5 years predicted by our XGBoost AFT and our ResNet. AF: Atrial Fibrillation, DM: Diabetes Mellitus, HTN: Hypertension, IHD: Ischemic Heart Disease, VHD: Valvular Heart Disease.}
	\label{fig:examples}
 
	\end{center}
\end{figure}

\section{Use Case: myHeartScore App}
\label{sec:myheartscore}

{Here, we provide a comprehensive understanding of how our research is applied in a real-world scenario to help prevent cardiovascular diseases such as HF. We present our app called \textit{myHeartScore} as a use case of our proposals. Besides, we detail the engineering and user interface elements that aid in assessing the cardiovascular risk of the users.}

\begin{figure}
    \centering
	\includegraphics[width=0.7\textwidth]{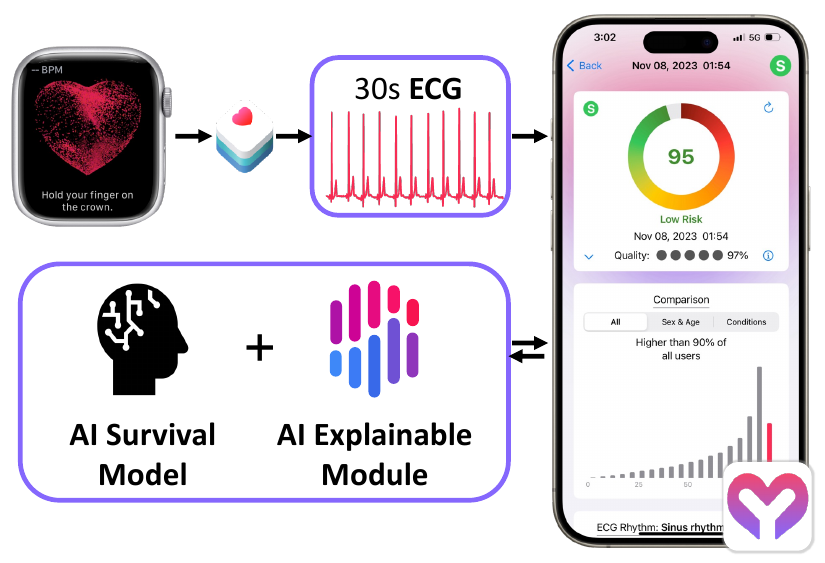}
    \caption{\color{black}Schematic of \textit{myHeartScore} leveraging the ECG from the Apple Watch}
	\label{fig:myheartscore}
\end{figure}

{\textit{MyHeartScore} is an iPhone app available on the App Store that helps uncover potential cardiovascular risks of the users \cite{myheartscore}. Its main feature is to provide users with a personalized health risk score based on their 30s ECG from Apple Watch and personal data. \textit{MyHeartScore}'s analysis relies on our HF survival models and two ECG features based on the rhythm of the ECG recording and the characteristic shape of the PQRST complex. Besides, it can be extended to consider the HRV statistics computed with the beat-to-beat measurement of the Apple Watch.} Figure \ref{fig:myheartscore} shows a schematic of the \textit{myHeartScore} application leveraging the Apple Watch and our ML survival model.

For this particular application, we chose our XGBoost model over ResNet. While our analyses in Section \ref{sec:experiments} showed that our ResNet model achieves better performance, XGBoost offers the advantages of a more compact architecture and a less resource-demanding inference, potentially enabling on-device predictions. Besides, its outputs are generally easier to interpret using explanability technologies \cite{ali2023explainable}. That is, the interpretations directly link to our more human-readable features of the ECG signal.

Upon the app's initial launch, users are prompted to provide their sex, age, and any pre-existing comorbidities. This information is later sent to our survival model along with the 30s ECG signal of their choice to output the corresponding survival curve. Given the estimated curve, a final risk score is provided as the cumulative probability of remaining without HF within five years. To enhance the understanding of the score, we have defined four levels of risk (low, moderate, mild, and severe) with the mean and standard deviation of the estimations for subjects with HF within five years and without HF with an observation time greater than five years in our database.

Beneath each computed score, we showcase two distinct ECG graphic visualizations. The first one exhibits the entire 30s ECG waveform, highlighting its RR intervals and rhythm features (average heart rate, SD1/SD2, and SDNN). The second visualization represents the ECG cycle template. Here, the user can visualize all cardiac cycles superimposed on each other and the timing and amplitudes of the PQRST complex. Besides, we display the negative or positive contributions of the two feature groups to a lower or higher risk score. These contributions are the sum of the SHAP values of each group's features. SHapley Additive exPlanations or SHAP \cite{NIPS2017_8a20a862} is a model agnostic explainability framework that permits to represent the prediction of a model as the sum of the expected model output and the feature contributions.

\begin{figure}
	\begin{center}
    \subfloat[Evaluated Score\label{fig:usecase_score}]{\fbox{\includegraphics[height=0.41\textwidth]{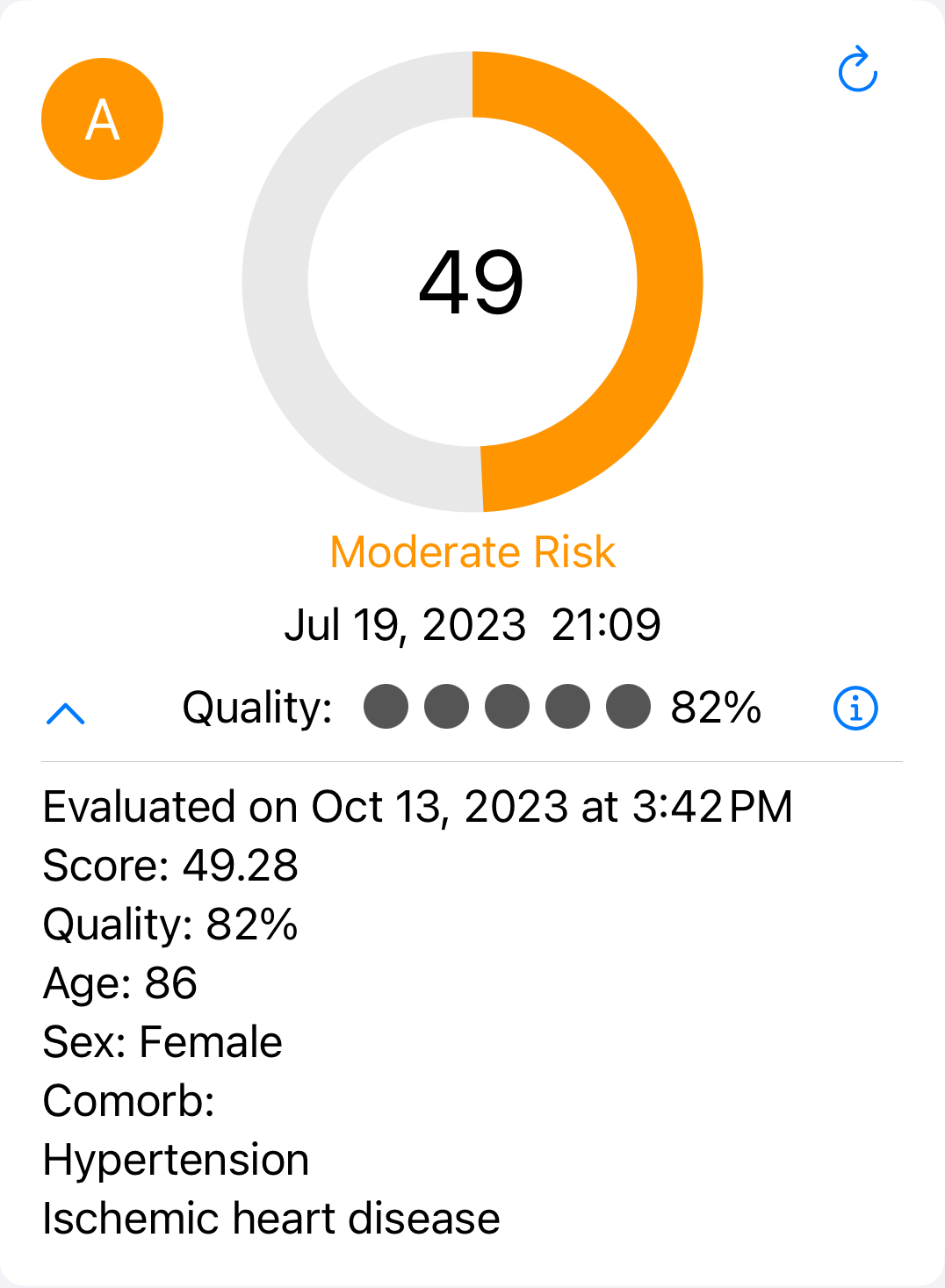}}}
    \subfloat[ECG Rhythm\label{fig:usecase_rhythm}]{\fbox{\includegraphics[height=0.41\textwidth]{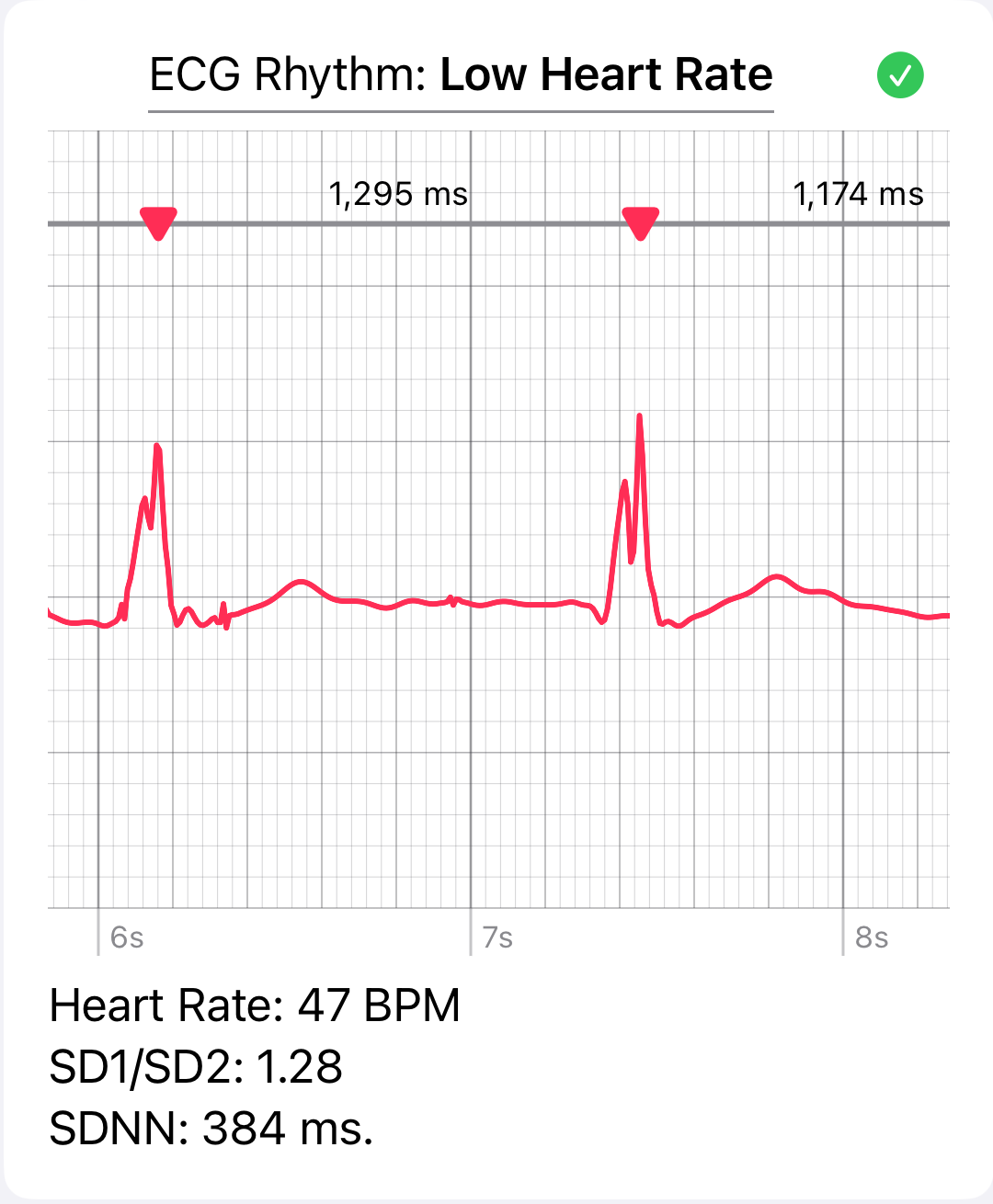}}}
    \subfloat[ECG Cycle Template\label{fig:usecase_shape}]{\fbox{\includegraphics[height=0.41\textwidth]{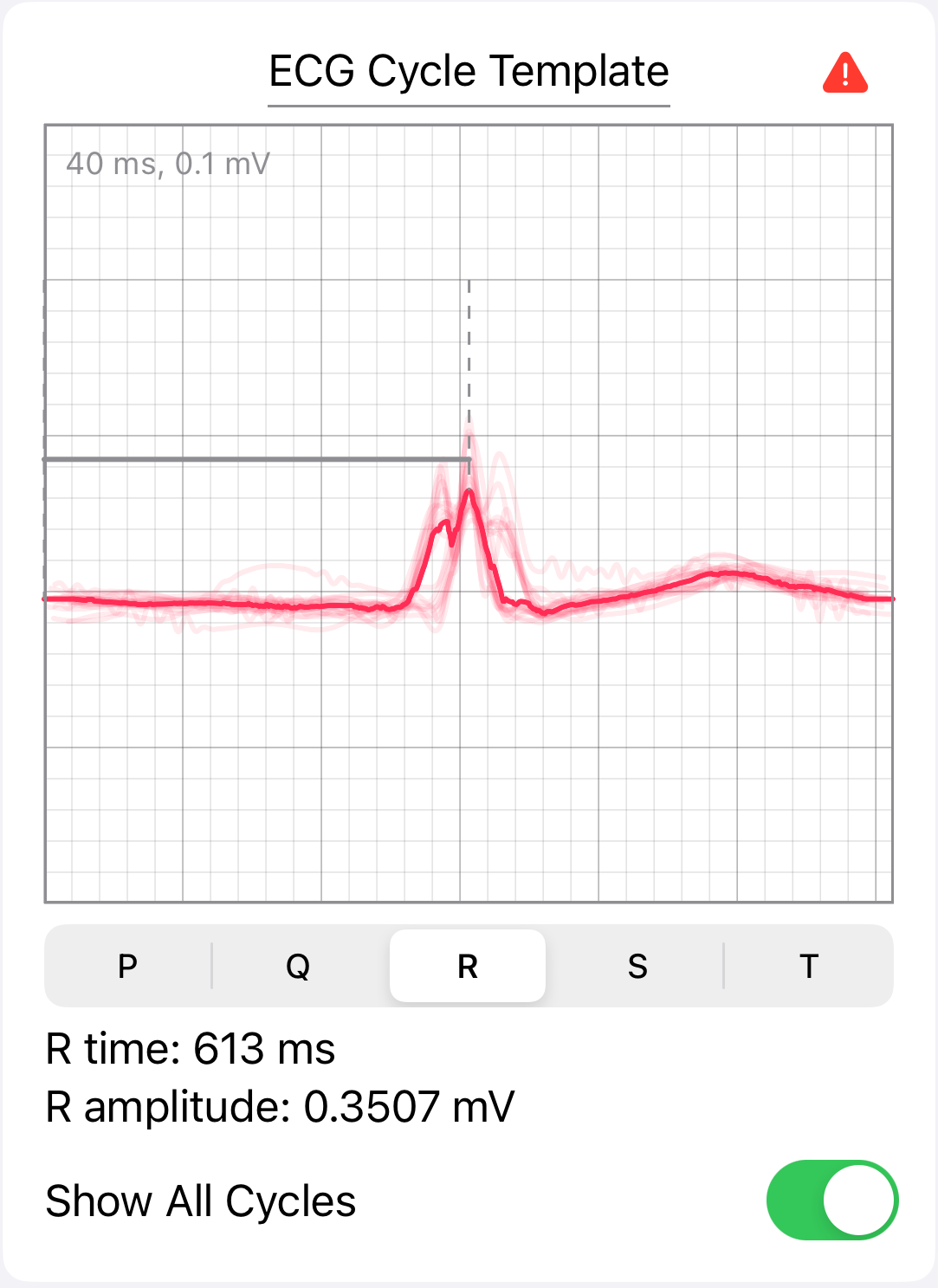}}}
    \caption{Screenshots of a ECG recording analyzed with myHeartScore app}
	\label{fig:usecases}		
	\end{center}
\end{figure}

Figure \ref{fig:usecases} shows an example of the above description for the \textit{myHeartScore} app. This example corresponds to the same volunteer and her recording exhibited in Figure \ref{fig:example0}. In Figure \ref{fig:usecase_score} the score and risk level are displayed together with her personal details. Figure \ref{fig:usecase_rhythm} and Figure \ref{fig:usecase_shape} present a scrollable visualization of the entire ECG and the plot of the cycle template, respectively. In the upper right of Figure \ref{fig:usecase_rhythm}, the green check icon indicates a slightly positive contribution from the ECG rhythm features. In contrast, the red warning symbol in Figure \ref{fig:usecase_shape} denotes a significant decrease in the score due to the PQRST complex features, which leads to the final score of 49.

\section{{Discussion and Conclusions}}
\label{sec:conclusions}

In this paper, we have presented a novel multi-modal approach for assessing the risk of HF hospitalization. This approach integrates short ECG recordings, sampled long-term HRV, demographic data, and patient medical history. Our approximate long-term HRV statistics were extracted by combining several short-term beat-to-beat measurements sampled during 24 hours, which leads to a better understanding of heart activity and improved model performance. {Furthermore, we have proposed two innovative survival frameworks: a feature-based ML model with comprehensive ECG and HRV features, whose backbone is XGBoost AFT, and a DL approach implemented with ResNet and a Transformer module specifically designed for learning risk factors directly from the ECG signal and HRV series.}

{Our multi-modal approach has demonstrated high performance for HF risk assessment compared to 14 different models and a competitive discriminatory capability across multiple open datasets for HF classification. We have also demonstrated that sampled long-term HRVs boost the model performance similarly to 24-hour exact HRVs.} Besides, we have shown that our approach can be effectively implemented in any wearable, promoting accessibility, convenience, and cost-effectiveness. The \textit{myHeartScore} App, utilizing Apple Watch ECG data and our ML survival models, showcases a practical application of our approach, allowing users to monitor their HF risk at any time.

In conclusion, our findings emphasize the importance of multi-modal learning and its potential to revolutionize HF risk assessment. By making HF risk assessment more accessible and convenient through wearable technology, we aim to contribute to its early detection and effective management, ultimately improving patient outcomes and reducing healthcare costs. This research lays the foundation for further exploration and implementation of multi-modal approaches in healthcare, demonstrating the potential for enhancing predictive modeling and risk assessment.

\appendix
\section{} 
\label{appendix}

{Table \ref{tab:res-models-app} shows the performance of 14 different models on the Holter ECG test set with and without approximate long-term \gls{HRV}s. We have grouped them into four categories: traditional survival models, feature-based models, signal-based or DL models, and multi-modal architectures. The last category includes those DL architectures that learn directly from the ECG signal and the HRV series. Therefore, they do not have a version without HRVs.}

\begin{table}[ht]

\resizebox{\textwidth}{!}{
\centering 
\begin{tabular}{l|rrrr|rrrr}
\toprule
{} &  \multicolumn{4}{c|}{\textbf{30s ECG}}  & \multicolumn{4}{c}{\textbf{Multi-modal (ECG \& HRVs)}} \\
\midrule
{} &  \textbf{C-index} &    \textbf{AUC} ($y\le5$) &     \textbf{iBS} &  \textbf{c/d AUC} &  \textbf{C-index} &    \textbf{AUC} ($y\le5$) &     \textbf{iBS} &  \textbf{c/d AUC} \\
\midrule
AFT              &   0.8009 &  0.8128 &  0.0664 &   0.8168 &   0.8234 &  0.8353 &  0.0646 &   0.8414 \\
CPH              &   0.8010 &  0.8129 &  0.0669 &   0.8166 &   0.8220 &  0.8337 &  0.0654 &   0.8390 \\
\midrule
MLP DeepHit      &   0.8025 &  0.8171 &  0.0691 &   0.8220 &   0.8284 &  0.8420 &  0.0709 &   0.8468 \\
RSF               &   0.8217 &  0.8369 &  \textbf{0.0632} &   0.8396 &   0.8330 &  0.8490 &  \textbf{0.0620} &   0.8543 \\
EST               &   0.8052 &  0.8235 &  0.0644 &   0.8237 &   0.8225 &  0.8347 &  0.0635 &   0.8388 \\
XGBoost AFT          &   0.8312 &  0.8440 &  0.0921 &   0.8508 &   0.8400 &  0.8517 &  0.0901 &   0.8598 \\
\midrule
ResCNN           &   0.8163 &  0.8307 &  0.0771 &   0.8318 &   0.8405 &  0.8547 &  0.0739 &   0.8613 \\
XceptionTime     &   0.8302 &  0.8429 &  0.0821 &   0.8479 &   0.8367 &  0.8482 &  0.0753 &   0.8566 \\
InceptionTime    &   0.8366 &  0.8490 &  0.0863 &   0.8556 &   0.8409 &  0.8545 &  0.0698 &   0.8613 \\
ResNet           &   \textbf{0.8403} &  \textbf{0.8545} &  0.0823 &   \textbf{0.8578} &   0.8468 &  0.8583 &  0.0685 &   0.8608 \\
\midrule
GRU-InceptionTime &   - &  - &  - &   - &   0.8441 &  0.8586 &  0.0734 &   0.8674 \\
GRU-ResNet       &   - &  - &  - &   - &   0.8469 &  0.8592 &  0.0767 &   0.8642 \\
TFM-InceptionTime &   - &  - &  - &   - &   0.8496 &  0.8649 &  0.0777 &   0.8690 \\
TFM-ResNet       &   - &  - &  - &   - &   \textbf{0.8537} &  \textbf{0.8688} &  0.0744 &   \textbf{0.8718} \\
\bottomrule
\end{tabular}
}
\caption{Performance on the Holter ECG test set with and without HRVs.}
\label{tab:res-models-app}
\end{table}

\bibliographystyle{model1-num-names}
\bibliography{reference.bib}

\end{document}